\documentclass[a4paper,11pt]{article}
\usepackage{jcappub}
\usepackage[T1]{fontenc}
\usepackage{subcaption}
\usepackage{graphicx}
\usepackage{amsmath}
\usepackage{amssymb}
\usepackage{lipsum}
\usepackage{comment}
\usepackage{multirow}
\usepackage{xcolor}
\usepackage{hyperref}
\usepackage[normalem]{ulem}
\pdfoutput=1

\usepackage{aas_macros}
\newcommand{\LCDM}{$\Lambda$CDM }

\newcommand{\Omm}{\Omega_\mathrm m}
\newcommand{\Omc}{\Omega_\mathrm c}
\newcommand{\Omb}{\Omega_\mathrm b}
\newcommand{\Omn}{\Omega_\nu}
\newcommand{\Mnu}{M_\nu}

\newcommand{\Pmm}{P_\mathrm{mm}}
\newcommand{\Pcc}{P_\mathrm{cc}}
\newcommand{\Pnn}{P_\mathrm{\nu\nu}}
\newcommand{\Pcn}{P_\mathrm{c\nu}}

\newcommand{\fc}{f_\mathrm{c}}
\newcommand{\fn}{f_\nu}

\newcommand{\Nbody}{$N$-body }

\newcommand{\react}{\texttt{ReACT}}
\newcommand{\baccoemu}{\texttt{baccoemu}}

\newcommand{\de}{\mathrm{d}}

\title{\boldmath{\fontsize{18.5}{11}\selectfont DEMNUni: comparing nonlinear power spectra prescriptions in the presence of massive neutrinos and dynamical dark energy}}
\author[a,b,c,d,e,f]{G. Parimbelli}
\emailAdd{gabriele.parimbelli@edu.unige.it}
\author[g]{, C. Carbone}
\author[h]{, J. Bel}
\author[i,j,k]{, B. Bose}
\author[l]{, M. Calabrese}
\author[m,g]{, E. Carella}
\author[n,o]{, M. Zennaro}
\affiliation[a]{Dipartimento di Fisica, Università degli studi di Genova,\\Via Dodecaneso 33, 16146 Genova, Italy}
\affiliation[b]{Università degli Studi Roma Tre,\\Via della Vasca Navale, 84, 00146 Rome, Italy}
\affiliation[c]{INFN -- Istituto Nazionale di Fisica Nucleare, Sezione di Roma Tre,\\Via della Vasca Navale, 84, 00146 Rome, Italy}
\affiliation[d]{INAF-OATs, Osservatorio Astronomico di Trieste,\\Via Tiepolo 11, 34131 Trieste, Italy}
\affiliation[e]{SISSA -- Scuola Internazionale Superiore di Studi Avanzati,\\Via Bonomea 265, 34136 Trieste, Italy}
\affiliation[f]{IFPU -- Institute for Fundamental Physics of the Universe,\\Via Beirut 2, 34151 Trieste, Italy}
\affiliation[g]{INAF -- Istituto di Astrofisica Spaziale e Fisica cosmica di Milano (IASF-MI), Via Alfonso Corti 12, I-20133 Milano, Italy}
\affiliation[h]{Aix Marseille Univ, Universit\'e de Toulon, CNRS, CPT, Marseille, France}
\affiliation[i]{D\'epartement de Physique Th\'eorique, Universit\'e de Gen\`eve, 24 quai Ernest Ansermet, 1211 Gen\`eve 4, Switzerland.}
\affiliation[j]{Institute for Computational Science, University of Zurich, Winterthurerstrasse 190, 8057 Zurich, Switzerland. } 
\affiliation[k]{Institute for Astronomy, University of Edinburgh, Royal Observatory, Blackford Hill, Edinburgh, EH9 3HJ, U.K.} 
\affiliation[l]{Astronomical Observatory of the Autonomous Region of the Aosta Valley (OAVdA), Loc. Lignan 39, I-11020, Nus (Aosta Valley), Italy}
\affiliation[m]{Dipartimento di Fisica ``Aldo Pontremoli'', Universit\`{a} degli Studi di Milano, via Celoria 16, I-20133 Milano, Italy}
\affiliation[n]{Donostia International Physics Center, Paseo Manuel de Lardizabal 4, 20018 Donostia-San Sebasti\'an, Gipuzkoa, Spain}
\affiliation[o]{Department of Physics (Astrophysics), University of Oxford, Denys Wilkinson Building, Keble Road, Oxford, OX1 3RH, United Kingdom.}
\abstract{
We provide an accurate comparison, against large cosmological \Nbody simulations, of different prescriptions for modelling nonlinear matter power spectra in the presence of massive neutrinos and dynamical dark energy.
We test the current most widely used approaches: fitting functions (HALOFIT and HMcode), the halo-model reaction (\react) and emulators (\baccoemu \ and EuclidEmulator2).
Focussing on redshifts $z\leq2$ and scales $k\lesssim 1 \ h/$Mpc (where the simulation mass resolution provides $\sim 1\%$ accuracy), we find that HMcode and \react \ considerably improve over the HALOFIT prescriptions of Smith and Takahashi (both combined with the Bird correction), with an overall agreement of 2\% for all the cosmological scenarios considered.
Concerning emulators, we find that, especially at low redshifts, EuclidEmulator2 remarkably agrees with the simulated spectra at $\lesssim 1\%$ level in scenarios with dynamical dark energy and massless neutrinos, reaching a maximum difference of $\sim 2\%$ at $z=2$. \baccoemu\ has a similar behaviour as EuclidEmulator2, except for a couple of dark energy models. In cosmologies with massive neutrinos, at $z=0$ all the nonlinear prescriptions improve their agreement with respect to the massless neutrino case, except for the Bird and TakaBird models which, however, are not tailored to $w_0$--$w_a$ models.
At $z>0$ we do not find a similar improvement when including massive neutrinos, probably due to the lower impact of neutrino free-streaming at higher redshifts; rather at $z=2$ EuclidEmulator2 exceeds $2\%$ agreement for some dark energy equation of state.
When considering ratios between the matter power spectrum computed in a given cosmological model and its \LCDM counterpart, all the tested prescriptions agree with simulated data, at sub-percent or percent level, depending on $z$.
Finally, we also test how nonlinear prescriptions compare against simulations when computing cosmic shear and angular galaxy clustering spectra.
For the former, we find a 2--3\% agreement for HMcode, \baccoemu, EuclidEmulator2 and \react; for the latter, due to the minimum stellar mass of the simulated galaxies, shot noise highly affects the signal and makes the discrepancies as high as 5\%.
}

\begin{document}
\maketitle

\flushbottom

%%%%%%%%%%%%%%%%%%%%%%%%%%%%%%%%%%%%%%%%%%%%%%%%%%%%
\section{Introduction}
\label{sec:intro}

The nonlinear power spectrum, $\Pmm(k)$, of matter density fluctuations, $\delta_\mathrm m \equiv \rho_\mathrm m/\bar\rho_\mathrm m -1$ (where $\rho_\mathrm m$ and $\bar\rho_\mathrm m$ represent the density field and background density of matter, respectively), is a quantity of utmost importance in the analysis of the large scale structure of the Universe.
Its modelling is essential to predict the great majority of cosmological observables. 
Current and upcoming galaxy surveys like Euclid\footnote{\url{https://www.euclid-ec.org/}}, the Vera C. Rubin Observatory\footnote{\url{https://www.lsst.org/}}, DES\footnote{\url{https://www.darkenergysurvey.org/}}, DESI\footnote{\url{https://desi.lbl.gov/}}, the Nancy Grace Roman Space Telescope\footnote{\url{https://roman.gsfc.nasa.gov/}} and SKA\footnote{\url{https://www.skatelescope.org/}} will measure galaxy clustering, cosmic shear, Lyman-$\alpha$ flux, HI intensity mapping and other relevant cosmological probes with unprecedented accuracy.
To this end, a high level of accuracy in the knowledge of the matter power spectrum down to very small scales ($k\sim 10 \ h/$Mpc) is required in order to optimally exploit high quality data whilst not biasing the constraints on the inferred cosmological parameters~\cite{Taylor:2018nrc,Impact_non-linear_recipe-Martinelli+21,Lacasa:2019flz}.

The computation of the matter power spectrum in the linear regime (i.e. for $\delta_\mathrm m \ll 1$) usually involves codes ({\it e.g.} \texttt{CAMB}\footnote{\url{https://camb.info/}}~\cite{CAMB} or \texttt{Class}\footnote{\url{https://lesgourg.github.io/class_public/class.html}}~\cite{CLASS}) that solve the Boltzmann equation satisfied by density perturbations.
These codes provide predictions accurate at the 0.1\% level in a computational time of about one second.
However, at small scales and late cosmic time the condition $\delta_\mathrm m \ll 1$ no longer applies, so that other methods must be invoked for the $\Pmm(k)$ computation.

On the one hand, nonlinear perturbation theory (PT), in all its versions ({\it e.g.} Standard Eulerian or Lagrangian PT~\cite{PT}, Renormalized PT~\cite{RPT}, Multipoint Propagator Theory~\cite{MPT-Bernardeau+12}, Effective Field Theory of large scale structure~\cite{EFT}), at low redshifts $z$ is accurate only at scales $k\lesssim 0.2 \ h/$Mpc at $z=0$ ~\cite{2009PhRvD..80d3531C,PT} and cannot capture the full cosmological information contained in the aforementioned observables.

On the other hand, \Nbody simulations provide a powerful tool to test gravity down to the deeply nonlinear regime of cosmological perturbations.
Regarding the nonlinear $\Pmm(k)$, different codes have been shown to agree, at $z=0$, at 1\% for $k<1 \ h/$Mpc and 2--3\% for $k<10 \ h/$Mpc, respectively~\cite{accuracy_power_spectrum-Schneider+16,GADGET_4}. 
However, cosmological simulations are computationally expensive and their direct use is prohibitive for parameter space sampling and cosmological inference, through {\it e.g.} Markov Chain Monte Carlo (MCMC) methods.
This problem can be circumvented through the use of fitting functions or emulators.
In both cases, a large suite of \Nbody simulations are first run with different cosmological parameters and for a large variety of cosmological models.
Then, in the former case, empirical formul\ae~can be fitted to the resulting nonlinear power spectra in order to minimize the residuals against simulated data~\cite{HALOFIT_Smith,HALOFIT_Bird,HALOFIT_Takahashi,HMcode2016_1,HMcode2016_2,HMcode2020,Smith-Angulo+19,Pk_equal-Casarini+16}.
In the latter case, an emulator is a regression model that learns the simulated matter power spectra (training set), interpolates them and provides accurate predictions for new models, i.e. cosmological parameters, not originally included in the training set ~\cite{baccoemu_sims,baccoemu_linear,baccoemu_baryons,EuclidEmulator,EuclidEmulator2,Coyote_I,Coyote_extended}.
These two methods immensely reduce the computational cost with respect to cosmological simulations at the expenses of a slightly larger uncertainty.

Most of these emulators and fitting functions were first built and tailored for the standard massless neutrino $\Lambda$ Cold Dark Matter ($\Lambda$CDM) model.
More recently a huge effort is being made to extend their application particularly to models with non-vanishing total neutrino mass $M_\nu = \sum_i m_{\nu,i}\neq 0$ and dynamical dark energy (DDE), i.e. dark energy with a redshift-dependent equation of state~\cite{chevallier2001accelerating,Linder2002}. However, the lack of reliable and precise theoretical predictions beyond the linear regime for such extended models still causes limitations in the constraining power on cosmological parameters \cite{DES_yr3_extension_to_LCDM}.

Alternatively, recent proposals have shown that it is possible to accurately model the nonlinear regime of structure formation using the halo-model~\cite{Halo_model,Mead:2016ybv,React_I,React_III,React_V}.
In particular, in the series of papers Ref.~\cite{React_I,React_II,React_III,React_IV,React_V,React_VI}, it has been shown that competitive accuracy can be achieved for a wide range of gravitational and cosmological models (including DDE and massive neutrinos) using the so called {\it halo-model reaction} (\react).
However, this approach relies on an accurate nonlinear prescription for the $\Lambda$CDM physics.

Extending these approaches to models that deviate from a constant $\Lambda$ is crucial for generalising data analysis and shed light on the nature of dark energy and ultimately the origin of the recent accelerated expansion of the Universe (see  {\it e.g.} Ref.~\cite{Huterer:2017buf} for a recent review).
This endeavour is fuelled by the fact that most galaxy surveys like Euclid and the Vera C. Rubin Observatory are designed precisely for tracing the evolution of the dark energy equation of state~\cite{Euclid_forecast+19,LSST_science,LSST_magnification_bias} and are expected to measure, for the first time, a non-zero total neutrino mass with high significance ($\gtrsim 2\sigma$)~\cite{neutrino_mass_Euclid-Chudakin+19}. 

With many ongoing surveys already taking data, and more upcoming in the next years, it is essential to understand which modelling prescriptions to adopt when analysing different probes, and what is their modelling uncertainty in order to infer unbiased cosmological constraints.
In this paper we compare the nonlinear matter power spectra predicted via the aforementioned approaches against the “Dark Energy and Massive Neutrino Universe” (\href{https://www.researchgate.net/project/DEMN-Universe-DEMNUni}{DEMNUni}) \Nbody simulations accounting for DDE in the background and massive neutrinos as a separate particle component.
We will also give a glimpse of how this affects primary observables like cosmic shear and angular galaxy clustering two point statistics.

This paper is organised as follows: in Sec.~\ref{sec:theory} we illustrate the effects of massive neutrinos and DDE on the matter power spectrum; in Sec.~\ref{sec:simulations} we describe the set of simulations used as reference data; in Sec.~\ref{sec:nonlinear_methods} we list all the nonlinear prescriptions considered in this work; in Sec.~\ref{sec:results} we show our main results; in Sec.~\ref{sec:conclusions} we draw our conclusions.

%%%%%%%%%%%%%%%%%%%%%%%%%%%%%%%%%%%%%%%%%%%%%%%%%%%%
\section{Massive neutrinos and dark energy}
\label{sec:theory}

The discovery of neutrino flavor oscillations (see {\it e.g.} Ref.~\cite{particle_physics_review} for a review) has confirmed that neutrinos have a nonzero mass.
On the one hand, particle physics experiments have placed a lower bound on the total neutrino mass, $M_\nu\gtrsim0.058$ eV ({\it e.g.} Ref.~\cite{massive_nu_cosmology+06}), and an upper bound of $M_\nu\lesssim 2.2-2.4$ eV~\cite{KATRIN2022}.
On the other hand, different analyses of cosmological data sets have provided upper limits of $M_\nu\lesssim 0.12-0.13$ eV at 95\% confidence level (CL) \cite{BOSS_Lymanalpha+15, Giusarma+16,neutrino_constraints-Vagnozzi+17,Planck2018,Abazajian2021,Neutrino_mass_constraint_Tanseri+22}, besides marginal preference for a non-null neutrino mass~\cite{Battye+14,Beutler+14,Pellejero-Ibanez2016,DiValentino+17}, and slight evidence for the so-called normal hierarchy~\cite{neutrino_hierarchy-Jimenez+22,Gariazzo2022,neutrino_hierarchy-Simpson+17}.
The most stringent constraint of $M_\nu\leq0.09$ eV at 95\% CL~\cite{most_stringent_neutrino_mass_constraint-DiValentino+22} has been inferred via the combination of Cosmic Microwave Background (CMB) temperature and polarization power spectra from Planck, CMB lensing, Supernovae Ia, Baryon Acoustic Oscillations (BAO) and redshift-space distortions from the eBOSS survey.

Cosmological probes, and the Large Scale Structure (LSS) in particular, have a high constraining power on neutrinos because of their large impact on the matter power spectrum both at linear and nonlinear levels~\cite{2018MNRAS.481.1486B,HALOFIT_Bird,2014JCAP...11..039B,HMcode2016_2,Mira_Titan_II,Tram:2018znz,neutrino_halo_model}.
Neutrinos decouple in the early Universe while being still relativistic but their high thermal velocities prevent them from clustering in regions smaller than the so-called free-streaming scale, $\lambda_\mathrm{fs}$ ({\it e.g.} Ref.~\cite{massive_nu_cosmology+06}).
For neutrino particles transitioning to the non-relativistic regime during the matter dominated era (as it is the case for active neutrinos), the comoving free-streaming wavenumber passes at the transition time through a minimum, $k_\mathrm{nr}\equiv k_\mathrm{fs}(z_\mathrm{nr})= 0.018 \ \Omm^{1/2} \left(M_\nu/1 \ \mathrm{eV}\right)^{1/2} \ h$/Mpc, $\Omm$ being the total matter background density (massive neutrino included) at $z=0$~\cite{massive_nu_cosmology+06}.
As a result, we expect a lower level of LSS clustering at wavenumbers $k> k_\mathrm{fs}(z_\mathrm{nr})$.

To be more quantitative, at redshifts relevant for galaxy clustering, the matter density fluctuation, $\delta_\mathrm m$, takes two contributions, one from cold dark matter plus baryons (CDM+b), and one from massive neutrinos:
\begin{equation}
    \delta_\mathrm m = \fc\delta_\mathrm c + \fn\delta_\nu,
    \label{eq:density_fluctuation_neutrino_cosmology}
\end{equation}

where the subscript ``c'' denotes CDM+b, $\fn$ is the neutrino fraction at $z=0$, that can be computed through
\begin{equation}
    \fn = \frac{\Mnu/(93.14 \ h^2 \ \mathrm{eV})}{\Omm},
    \label{eq:neutrino_fraction}
\end{equation}

and $\fc = 1-\fn$.
The total matter power spectrum can be computed by taking the ensemble average of the square modulus in Fourier space of Eq.~\eqref{eq:density_fluctuation_neutrino_cosmology}
\begin{equation}
    \Pmm(k) = \fc^2\Pcc(k)+2\fc\fn\Pcn(k)+\fn^2\Pnn(k),
    \label{eq:power_spectrum_cdm_nu}
\end{equation}

where we have decomposed $\Pmm$ in its contributions from CDM+b ($\Pcc$), neutrinos ($\Pnn$) and their cross-correlation ($\Pcn$).

DDE is a phenomenological attempt to detect deviations from the cosmological constant $\Lambda$ by means of a Taylor expansion of the Equation of State (EoS) of the associated DE fluid, parametrised à la Chevallier-Polarski-Linder (CPL) \cite{chevallier2001accelerating, Linder2002}. 
\begin{equation}
    w(a) = w_0 + (1-a) \ w_a.
    \label{eq:dark_energy_state_parameter}
\end{equation}

Current constraints from Planck+SNe+BAO yield $w_0=-0.957\pm0.080$ and $w_a=-0.29^{+0.32}_{-0.26}$ \cite{Planck2018}.
Upcoming surveys are expected to largely improve the uncertainties on these parameters, with $\sigma(w_0)\approx 1\%$ and $\sigma(w_a)\approx 10\%$~\cite{Euclid_forecast+19,LSST_science}.

\begin{figure}[t]
\centering 
\includegraphics[width=0.99\textwidth]{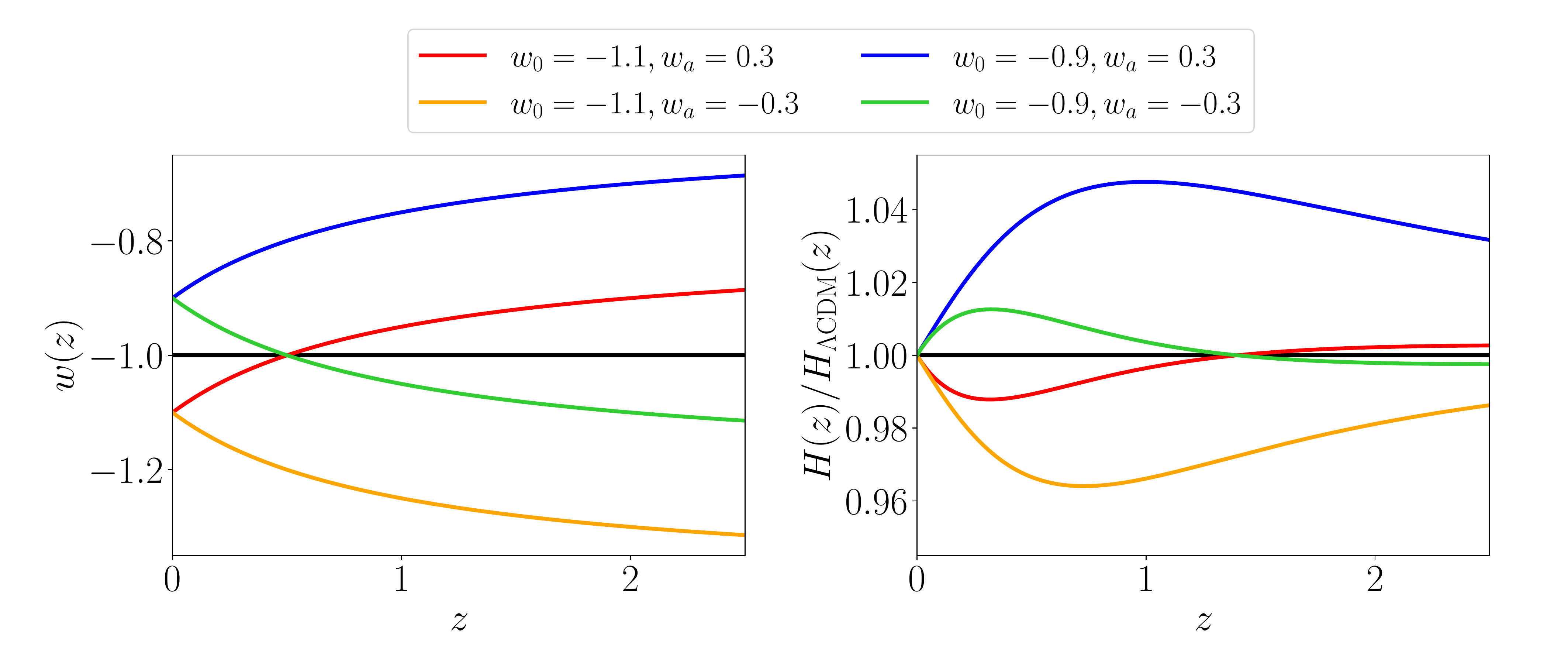}
\caption{Evolution of the EoS parameters of dynamical dark energy (left) and its impact on the Hubble parameter (right) for the cosmological models used in this work. The black lines represent the \LCDM model of reference; the coloured lines represent DDE models: red for ($w_0=-1.1$, $w_a=0.3$), orange for ($w_0=-1.1$, $w_a=-0.3$), blue for ($w_0=-0.9$, $w_a=0.3$) and green for ($w_0=-0.9$, $w_a=-0.3$).}
\label{fig:effect_on_Hz}
\end{figure}

\begin{figure}[t]
\centering 
\includegraphics[width=0.99\textwidth]{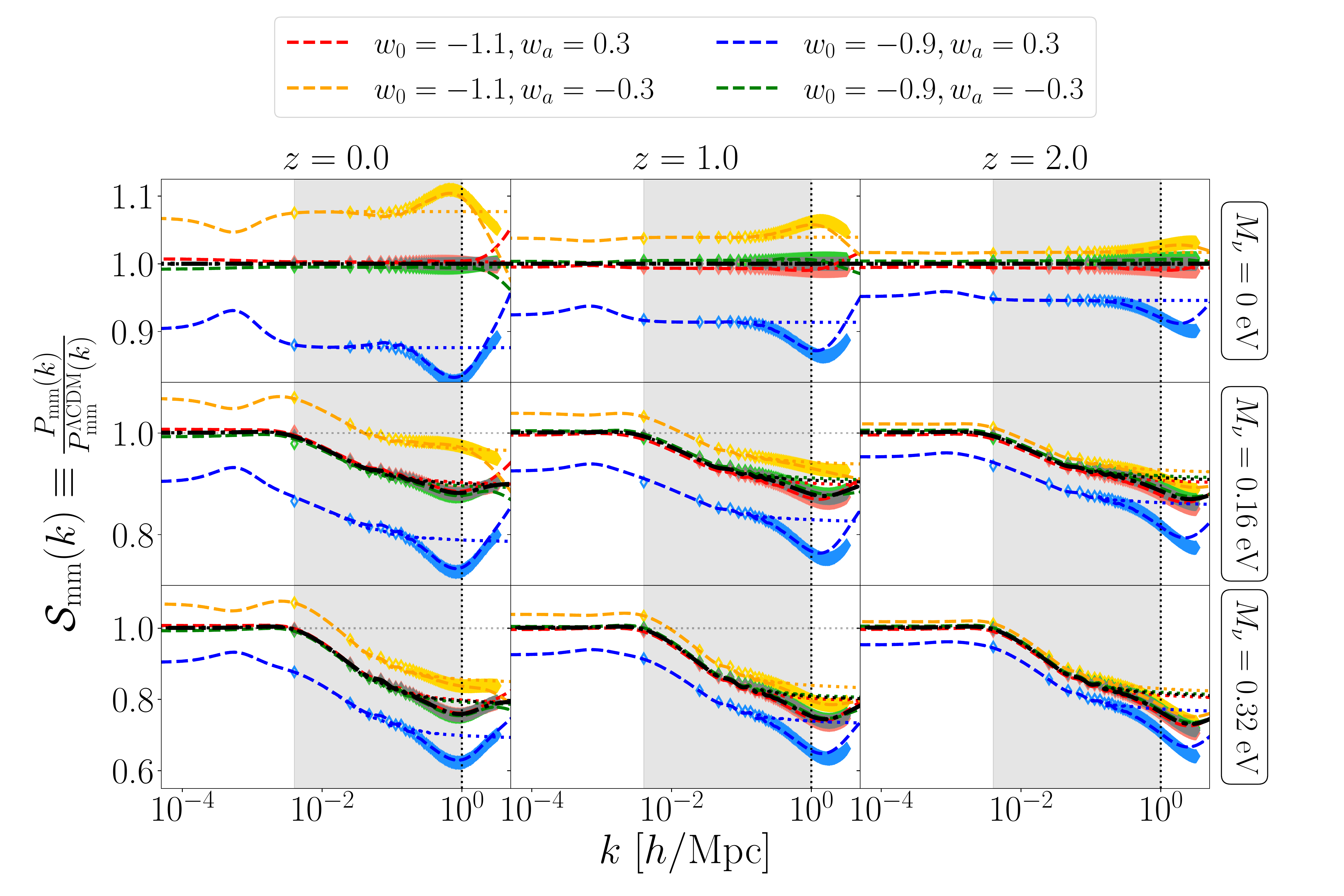}
\caption{Response $\mathcal S_{\rm mm}(k)$, Eq.~\eqref{eq:response_neutrino_dark_energy}, of $\Pmm(k)$ to the introduction of massive neutrinos and dynamical dark energy at $z=0$ (left), $z=1$ (middle) and $z=2$ (right).
We show the cosmological models described in the text (see Sec.~\ref{sec:simulations} and Tab.~\ref{tab:simulation_sets_table}).
Each panel contains the prediction for five different cosmologies: the top row contains the simulations with vanishing neutrino mass, while the middle and bottom rows contain simulations with $M_\nu=0.16, 0.32$ eV, respectively.
Dotted lines represent $\mathcal S_{\rm mm}(k)$ of the linear matter power spectrum, $\Pmm^{\rm L}(k)$, computed with CAMB, while dashed lines represent $\mathcal S_{\rm mm}(k)$ for the nonlinear $\Pmm^{\rm NL}(k)$ computed using the HMcode2020 prescription.
The black dashed line is the response when a nonzero $M_\nu$ (displayed on the right of the plot) is introduced.
The coloured lines represent the responses for models with different dark energy EoS:
red for ($w_0=-1.1$, $w_a=0.3$), orange for ($w_0=-1.1$, $w_a=-0.3$), blue for ($w_0=-0.9$, $w_a=0.3)$ and green for $(w_0=-0.9$, $w_a=-0.3$).
The fainter coloured diamonds are corresponding measurements from the DEMNUni simulations (see Sec. \ref{sec:simulations}).
The vertical grey shaded bands represent the range of scales considered in this work.
}
\label{fig:effect_on_Pk}
\end{figure}

Here we assume that DDE does not cluster, but affects only the background evolution of the Universe and, therefore, the linear growth of LSS via the expansion rate, $H(z)$~\cite{Cosmological_constant-Carroll+92}. DDE introduces a significant effect on the matter power spectrum by means of its impact on the background expansion.
In Fig.~\ref{fig:effect_on_Hz} we show the redshift evolution of the dark energy EoS parameters of the DEMNUni set (left) and their effect on $H(z)$ with respect to the \LCDM model (right).
Deviations in the background expansion rate can get as large as $4-5\%$ at $z\sim 1$ for $(w_0=-0.9, w_a=0.3)$ and $(w_0=-1.1, w_a=-0.3)$, i.e. for the models whose dark energy EoS deviates most from $w_0=-1$.
At $z=0$ the different values of $H(z)$ match by construction the \LCDM case, while they also slowly converge to the same value at large redshifts, where the Universe becomes matter dominated and DDE plays a more marginal role in the background evolution.
%---------------------------------------------------
%---------------------------------------------------
To quantify the impact of massive neutrinos and DDE on the \textit{total} matter power spectrum, we define a response function as the ratio between $\Pmm(k)$ in a given cosmology and $\Pmm(k)$ in a \LCDM cosmology having the same total matter and baryon densities, Hubble constant, scalar spectral index and amplitude of primordial curvature perturbations:
\begin{equation}
    \mathcal S_\mathrm{mm}(k) = \frac{\Pmm(k)}{\Pmm^{\Lambda\mathrm{CDM}}(k)}.
    \label{eq:response_neutrino_dark_energy}
\end{equation}
In Fig.~\ref{fig:effect_on_Pk} we show the response functions at $z=0$ (left panels), at $z=1$ (middle panels) and at $z=2$ (right panels).
The lines in each panel represent the linear and nonlinear $\mathcal S(k)$, respectively, due to the neutrino mass and the dark energy equation of state, with the same colour scheme used in Fig. \ref{fig:effect_on_Hz}.
Since the plot extends to very large scales, which are well outside the horizon, we specify that, in Fig.~\ref{fig:effect_on_Pk},
all the linear spectra have been computed using the CAMB code in the Newtonian gauge, and the theoretical nonlinear spectra with the HMcode2020 formul\ae~(see Ref. \cite{HMcode2020} and Sec.~\ref{sec:nonlinear_methods}).
The linear regime is denoted with a dotted line and the small-scale suppression of $\Pmm(k)$ and $\Pcc(k)$ due to the neutrino presence can be quantified respectively as~\cite{massive_nu_cosmology+06,DEMNUni_simulations_2}
\begin{equation}
    \mathcal S_\mathrm{mm}^\mathrm{L}(k) \approx 1-8\fn
    \qquad
    \mathcal S_\mathrm{cc}^\mathrm{L}(k) \approx 1-6\fn \, .
\end{equation}
In the nonlinear regime (dashed lines) the neutrino-induced suppression takes the well-known ``spoon-like'' shape.
The maximum depth of the suppression is $\mathcal S_\mathrm{mm}^\mathrm{NL}\approx 1-10\fn$, with a turnaround at scales dominated by the so-called 1-halo term, reflecting the fact that the number of small halos contributing to the dominant 1-halo term on scales $k>1 \ h$/Mpc beyond the turnaround, is negligibly affected by the presence of massive neutrinos~\cite{neutrino_halo_model}.

When using Eq.~\eqref{eq:power_spectrum_cdm_nu} in the nonlinear regime, one should in principle use the nonlinear spectra of all the components.
However, it was shown in Ref.~\cite{DEMNUni_simulations_2} that the nonlinear contributions by $\Pcn$ and $\Pnn$ are negligible with respect to $\Pcc$. Therefore, the theoretical nonlinear total matter power spectrum, $\Pmm^\mathrm{NL}(k)$, in cosmologies with massive neutrinos can be computed, within $1\%$ accuracy on the scales considered in this work, by using the linear $\Pcn^\mathrm{L}(k)$ and $\Pnn^\mathrm{L}(k)$, and the nonlinear $\Pcc^\mathrm{NL}(k)$ in Eq.~\eqref{eq:power_spectrum_cdm_nu}. Hereafter, we will adopt this prescription.
%---------------------------------------------------
%---------------------------------------------------

In Fig.~\ref{fig:effect_on_Pk} the coloured dashed lines represent models with DDE, and the top panels show results for the massless neutrino case.
Here, it is possible to observe a shift in the $\Pmm$ amplitude which is basically scale-independent at wavenumbers, $k$, larger than the horizon scale (which at $z=0$ is of order $10^{-4}-10^{-3} \ h/$Mpc) and smaller than scales, $k\sim 0.1 \ h/$Mpc, where nonlinear effects start to become important.
If we think to the halo-model, at these scales the scale-dependent shape is probably due to the fact that, between the $\Lambda$CDM and the considered cosmology, the fraction of big halos is very different while the fraction of small halos is very similar.
In particular, depending on the dark energy EoS, this translates into bumps (for $w_0=-1.1$) and wells (for $w_0=-0.9$) whose depth increases and position decreases with cosmic time (see the differences between the left and right panels of Fig.~\ref{fig:effect_on_Pk}).
Although we do not show it, the response function, $\mathcal S_{\nu+\mathrm{DDE}}$, for the combination of massive neutrinos and DDE is remarkably close to the product, $\mathcal S_{\nu} \times \mathcal S_\mathrm{DDE}$, of the two response functions alone: the agreement is at percent level at $z=0$, and as expected it gets even better at higher redshifts, when the dark energy density parameter and the effect of DDE on the matter power spectrum become smaller.
Finally, we notice how the two models with $w_0=-1.1, w_a=0.3$ (red lines) and $w_0=-0.9,w_a=-0.3$ (cyan line) are very close to each other and in turn to the \LCDM prediction at linear scales, confirming the same degeneracy pattern also visible in CMB analyses (see {\it e.g.} Fig.~30 of Ref. \cite{Planck2018}).

%%%%%%%%%%%%%%%%%%%%%%%%%%%%%%%%%%%%%%%%%%%%%%%%%%%%
\section{The DEMNUni simulations}
\label{sec:simulations}

Our goal is to compare different predictions of the nonlinear matter power spectrum against \Nbody simulations.
To this end, we make use of the “Dark Energy and Massive Neutrino Universe” (\href{https://www.researchgate.net/project/DEMN-Universe-DEMNUni}{DEMNUni}) suite~\cite{DEMNUni_simulations}.

The DEMNUni simulations have been produced with the aim of investigating the clustering of large scale structures in the presence of massive neutrinos and DDE and they were conceived for the nonlinear analysis and modelling of different probes, including dark matter halo- and galaxy-clustering~\cite{DEMNUni_simulations_2,Moresco2017,Zennaro2018,Ruggeri2018,Bel2019,Parimbelli2021, SHAM-Carella_in_prep}, CMB lensing, SZ and ISW effects~\cite{Roncarelli2015,DEMNUni_simulations,fabbian2018}, cosmic void statistics~\cite{Kreisch2019,Schuster2019,Verza2019, Verza2022}, and cross-correlations among these probes~\cite{Vielzeuf2022_inprep, Cuozzo2022_inprep}.

To this end, they combine a good mass resolution with a large volume to include perturbations both at large and small scales.
In fact, these simulations follow the evolution of 2048$^3$ CDM and, when present, 2048$^3$ neutrino particles in a box of side $L=2 \ \mathrm{Gpc}/h$.
The fundamental frequency of the comoving particle snapshot is therefore $k_\mathrm F \approx 3\times 10^{-3} \ h/$Mpc.

They were performed using the tree particle mesh-smoothed particle hydrodynamics (TreePM-SPH) code GADGET-3, an improved version of the code described in Ref.~\cite{GADGET_2}, specifically modified in Ref.~\cite{GADGET_3} to account for the presence of massive neutrinos.
This particular version of the code follows the evolution of CDM and neutrino particles as two separate collisionless fluids.
To save computational time, however, the calculation of the short-range tree force induced by the neutrino component can be neglected at early times.
Indeed, given the low mass and the consequent high velocity dispersion, neutrinos have a clustering scale which is much larger than the CDM one.
This results in a different scale resolution for the two components, which for neutrinos is fixed by the PM grid (chosen with a number of cells eight times larger than the number of particles), while for CDM particles is larger and given by the tree-force (for more details see Ref.~\cite{GADGET_3}).
As shown in Ref.~\cite{Non-linear_evo_neutrinos-Paco+12}, the application of the short-range tree force is required only when dealing with neutrino density profiles inside massive halos at low redshifts.
Therefore the choice of neglecting it at high redshifts does not affect the scales in which we are interested ($k\lesssim 1 \ h/$Mpc).

The reference cosmological parameters are chosen to be a baseline Planck13 cosmology~\cite{Planck2013}, with massless neutrinos and $\Omm = \Omc+\Omb+\Omega_\nu = 0.32, \ \Omb=0.05, \ h=0.67, \ n_\mathrm s=0.96, \ A_\mathrm s = 2.1265\times 10^{-9}$.
Given these values, the reference (i.e. in the massless neutrino case) CDM mass resolution is $m^p_\mathrm c = 8.27\times 10^{10} \ M_\odot/h$ and is decreased according to the mass of neutrino particles, in order to keep the same $\Omm$ among all the DEMNUni simulations.
The softening length is 20 kpc/$h$.
Massive neutrinos are assumed to come in three mass-degenerate species.
The sum of their masses is varied over the values $M_\nu=0, 0.16, 0.32$ eV.
To keep $\Omm$ fixed, an increase in $\Omn$ yields a decrease in $\Omc$.
For each value of $M_\nu$, five different simulations were run: one with ($w_0=-1$, $w_a=0$) and four with various combinations of $w_0=(-0.9, -1.1)$ and $w_a=(-0.3, 0.3)$, for a total of 15 simulations, summarised in Tab.~\ref{tab:simulation_sets_table}.

The simulations are initialized at $z_\mathrm{ini}=99$ with Zel'dovich initial conditions.
The initial power spectrum is rescaled to the initial redshift via the rescaling method developed in Ref.~\cite{rescaling_IC-Zennaro+17}.
Initial conditions are then generated with a modified version of the \texttt{N-GenIC} software, assuming Rayleigh random amplitudes and uniform random phases. For each simulation, 63 snapshots, logarithmically equispaced in the scale factor $a$, are saved. Moreover, about 400 TB of data in particle comoving snapshots, halo and galaxy catalogues, projected density maps, and power spectra of different particle species are stored and available upon request.
%---------------------------------------------------
\begin{figure}[t]
\centering 
\includegraphics[width=0.99\textwidth]{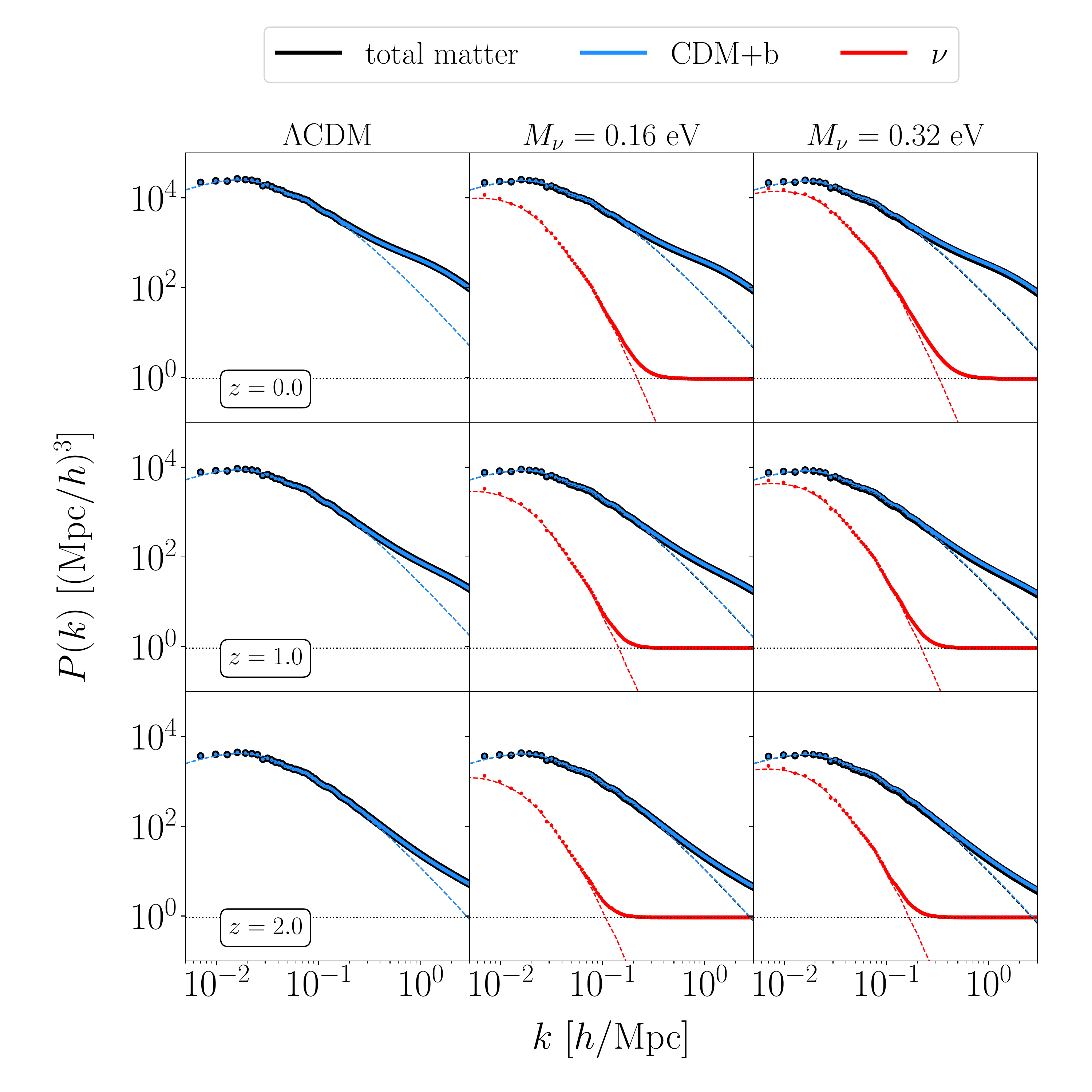}
\caption{Power spectra measured from three runs of the DEMNUni simulations presented here:  \LCDM (left column), ($w_0=-1$,$w_a=0$) and $M_\nu=0.16$ eV (middle column), ($w_0=-1$,$w_a=0$)  and $M_\nu=0.32$ eV (right column). Different rows label redshift $z=0,1,2$, respectively. Black, blue and red dots label measured total matter, CDM+b and neutrino power spectra, while dashed lines of the same colour refer to the corresponding linear spectra at the same redshift. The dotted horizontal line represents the Poisson shot noise for CDM+b and neutrinos.}
\label{fig:DEMNUni_Pk}
\end{figure}
%---------------------------------------------------

In this work we consider five particle snapshots at redshifts $z=2.0, 1.5, 1, 0.5, 0$. We compute CDM, neutrino and total matter auto-power spectra on a $2048^3$ regular grid with a Piecewise Continuous Spline mass assignment scheme and employ the interlacing technique as described in Ref.~\cite{sefusatti2016}. With this setting we expect a reliable estimate of the power spectra up to the Nyquist frequency $k_\mathrm{Nyq} \approx 3 \ h/$Mpc.
However, given the mass resolution, we expect the DEMNUni simulations to be accurate within $\sim 1\%$ percent level up to $k\sim 1 \ h/$Mpc~\cite{accuracy_power_spectrum-Schneider+16}.
We mention here that in Ref. \cite{baccoemu_sims} an updated comparison among different \Nbody codes was performed.
Using an empirical correction depending on the softening length, they were able to extend the expected accuracy of simulated power spectra beyond the nominal accuracy.
However, to be more conservative and given the value of the Nyquist frequency of the measurements, we will keep at $1 \ h/$Mpc the benchmark maximum scale for 1\% accuracy of the DEMNUni power spectra.

In Figs.~\ref{fig:effect_on_Pk}--\ref{fig:DEMNUni_Pk} we present examples of measured power spectra from the DEMNUni simulations. In particular, on the one hand, Fig.~\ref{fig:DEMNUni_Pk} shows the total matter, CDM+b and neutrino power spectra, together with their linear predictions, at redshifts $z=0,1,2$, in three cosmological scenarios: \LCDM (massless neutrino case); ($w_0=-1$,$w_a=0$) and $M_\nu=0.16$ eV; ($w_0=-1$,$w_a=0$)  and $M_\nu=0.32$ eV. On the other hand, the coloured diamonds in Fig.~\ref{fig:effect_on_Pk} show the measured responses, with respect to the \LCDM case, of the matter power spectra in all the cosmological scenarios covered by the DEMNUni suite.
%---------------
\begin{table}
	\centering
	\begin{tabular}{|c|c|c|c|c|c|c|}
	    \hline 
		\textbf{No.}  & $M_\nu$ (eV) & $\Omc+\Omb$ & $w_0$ & $w_a$ & $m^p_\mathrm c \ [M_\odot/h]$ & $m^p_\nu \ [M_\odot/h]$ \\
		\hline 
		\hline
		    \textbf{1}  & \multirow{5}{*}{0}   & \multirow{5}{*}{0.3200} & --1   &    0   & \multirow{5}{*}{$8.27\times 10^{10}$} & \multirow{5}{*}{$-$} \\ \cline{1-1} \cline{4-5}
	        \textbf{2}  &  &  & --0.9 &  --0.3 &  &  \\ \cline{1-1} \cline{4-5}
	        \textbf{3}  &  &  & --0.9 & \  0.3 &  &  \\ \cline{1-1} \cline{4-5}
	        \textbf{4}  &  &  & --1.1 &  --0.3 &  &  \\ \cline{1-1} \cline{4-5}
	        \textbf{5}  &  &  & --1.1 & \  0.3 &  &  \\ \hline\hline
	        \textbf{6}  & \multirow{5}{*}{0.16} & \multirow{5}{*}{0.3162} & --1   &   0   & \multirow{5}{*}{$8.17\times 10^{10}$} & \multirow{5}{*}{$9.97\times 10^{8}$} \\ \cline{1-1} \cline{4-5}
	        \textbf{7}  &  &  & --0.9 &  --0.3 &  &  \\ \cline{1-1} \cline{4-5}
	        \textbf{8}  &  &  & --0.9 & \  0.3 &  &  \\ \cline{1-1} \cline{4-5}
	        \textbf{9}  &  &  & --1.1 &  --0.3 &  &  \\ \cline{1-1} \cline{4-5}
	        \textbf{10} &  &  & --1.1 & \  0.3 &  &  \\ \hline\hline
	        \textbf{11} & \multirow{5}{*}{0.32} & \multirow{5}{*}{0.3123} & --1   &  0   & \multirow{5}{*}{$8.07\times 10^{10}$} & \multirow{5}{*}{$1.99\times 10^{9}$} \\ \cline{1-1} \cline{4-5}
	        \textbf{12} &  &  & --0.9 &  --0.3 &  &  \\ \cline{1-1} \cline{4-5}
	        \textbf{13} &  &  & --0.9 & \  0.3 &  &  \\ \cline{1-1} \cline{4-5}
	        \textbf{14} &  &  & --1.1 &  --0.3 &  &  \\ \cline{1-1} \cline{4-5}
	        \textbf{15} &  &  & --1.1 & \  0.3 &  &  \\ \hline
	\end{tabular}
	\caption{Specifications of the DEMNUni simulations used in this work. First column: simulation number; second column: sum of neutrino masses; third column: CDM and baryons density parameters; fourth and fifth columns: dark energy EoS parameters; sixth column: mass of CDM particles in the simulations; seventh column: mass of $\nu$ particles in the simulations.}
	\label{tab:simulation_sets_table}
\end{table}
%---------------
%%%%%%%%%%%%%%%%%%%%%%%%%%%%%%%%%%%%%%%%%%%%%%%%%%%%
\section{Modelling the nonlinear matter power spectrum}
\label{sec:nonlinear_methods}

We briefly describe in this Sec. the nonlinear prescriptions considered in this work for the modelling of $\Pmm^{\rm NL}$.
In general, we can recognise three different categories to which they belong: fitting functions, halo-models and emulators.
We quickly review their main features and summarise them in Tab.~\ref{tab:nonlinear_models}.
%---------------
\begin{table}
	\centering
	\resizebox{\textwidth}{!}{
	\begin{tabular}{|c|c|c|c|c|}
	    \hline 
		\textbf{Name}  & \textbf{Ref.} & \textbf{Type} & \textbf{$N_\mathrm{par/nodes}$} & \textbf{Validation} \\
		\hline 
		\hline
		    HALOFIT+Bird & \cite{HALOFIT_Smith,HALOFIT_Bird} & FF & 30+6 & \Nbody simulations (HYDRA) \\
		\hline
	        Takahashi+Bird & \cite{HALOFIT_Takahashi} & FF & 34+6 & \Nbody simulations (Gadget2) \\
	    \hline
	        HMcode2016 & \cite{HMcode2016_1,HMcode2016_2} & FF/HM & 12+2 & \textsc{Cosmic emu} +  Ref. \cite{neutrino_halo_model} \\
	    \hline
	        HMcode2020 & \cite{HMcode2020} & FF/HM & 12 & \textsc{Franken emu/Mira Titan} + Ref. \cite{neutrino_halo_model} \\
	    \hline
	        ReACT & \cite{React_I,React_II,React_III,React_IV,React_V,React_VI} & HM & None$^*$ & \Nbody and hydro simulations/\texttt{baccoemu} and EuclidEmulator2 \\
	    \hline
	        baccoemu & \cite{baccoemu_sims,baccoemu_linear,baccoemu_baryons} & E & 800 & BACCO simulations (L-Gadget3)  \\
	    \hline
	        EuclidEmulator2 & \cite{EuclidEmulator,EuclidEmulator2} & E & 108 & simulations (PKDGRAV3) \\
	    \hline
	\end{tabular}
	}
	\caption{List of the nonlinear prescriptions of $\Pmm^{\rm NL}$ analysed in this work. In the third column, the acronyms FF, HM and E stand for ``fitting function'', ``halo-model'' and ``emulator'', respectively. In the fourth column we indicate the number of free parameters (for fitting functions) or nodes (for emulators) used to fit the power spectra or to train the emulator. ${}^*$\react, despite having no free parameters itself, inherits that number from the method used to predict the pseudo spectrum.}
	\label{tab:nonlinear_models}
\end{table}
%---------------

One of the first significant attempts to accurately capture the nonlinear clustering of matter through a fitting function was the HALOFIT model by Smith \textit{et al.}~\cite{HALOFIT_Smith}.
In this model the matter power spectrum is the sum of two terms, a \textit{quasi-linear} one that reflects large scale matter density fluctuations and a \textit{halo} term that becomes dominant at large $k$ and describes the clustering of dark matter (DM) inside collapsed structures.
Despite the name, HALOFIT can be categorised as a fitting function, rather than a proper halo model. Indeed, while HALOFIT and the halo model share some features, like the splitting of the power spectrum into a quasi-linear and a nonlinear terms and the idea that the bulk of the cosmology dependence is captured by the mass density fluctuation function $\sigma(M)$, the latter naturally includes further ingredients such as spherical collapse, halo mass functions and the mass-concentrations relation.
The limited resolution of the \Nbody simulations against which it was tested, prevented from obtaining a high accuracy modelling in the deep nonlinear regime.
A global re-fitting of the free parameters against high-resolution simulations was later performed by Takahashi \textit{et al.}~\cite{HALOFIT_Takahashi}, together with an extension to dark energy with $w_0=$ constant models.
The revised formula provides an accurate description of the matter clustering with a precision of 5\% for $k\leq1 \ h/$Mpc at $z\leq10$, and 10\% for $1<k/(h/\mathrm{Mpc})\leq10$ at $z\leq3$.
Originally not designed for massive neutrino cosmologies, both the Smith and Takahashi prescriptions have been extended to such models by including the correction by Ref.~\cite{HALOFIT_Bird}. We refer to these as the ``Bird'' and ``TakaBird`` prescriptions, respectively.

HMcode, in its two versions of 2016~\cite{HMcode2016_1,HMcode2016_2} and 2020~\cite{HMcode2020}, lies at the frontier between fitting functions and the halo-model.
The basic idea is that the halo-model (see {\it e.g.} Ref.~\cite{Halo_model}) can describe the broad-band shape of the power spectrum but due to its simplistic assumptions cannot provide an accurate description of the details.
A new set of effective parameters were thus introduced to alleviate these discrepancies.
These parameters were fitted against the \textsc{Cosmic emu} emulator~\cite{Coyote_I,Coyote_extended} for HMcode2016 and the \textsc{Franken emu}~\cite{Coyote_extended} emulator for the HMcode2020.
Massive neutrinos are incorporated in accordance with the ``CDM prescription'', which helps recover the universality of the halo mass function and a scale-independent halo bias at the largest scales~\cite{cosmo_nu_1,cosmo_nu_2,cosmo_nu_3,cosmo_nu-Ichiki,Bias_neutrinos_CDM_prescription-Vagnozzi+18,Bias_neutrinos-Raccanelli+22}.
In other words, all relevant quantities are obtained replacing the linear total matter power spectrum, $\Pmm^{\rm L}$, with the CDM+b one, $\Pcc^{\rm L}$.
\textsc{Mira Titan}~\cite{Mira_Titan_I,Mira_Titan_II} nodes, containing massive neutrinos and DDE extensions, were naturally recovered by using HMcode2016 in combination with fitting functions for spherical collapse density thresholds ($\delta_\mathrm c$ and $\Delta_\mathrm{vir}$) in presence of massive neutrinos, fitted to simulations performed in Ref. \cite{neutrino_halo_model}.
The quoted overall precision for HMcode2016 is 5\% for $z\leq2$ and $k\leq 10 \ h/$Mpc. This value is comparable to the precision of \textsc{Cosmic emu} itself, so that HMcode2016 was as accurate as possible at the time the Coyote simulations were run.
HMcode2020, which also includes an improved BAO damping prescription, has a quoted overall RMS precision is quoted to be 2.5\% over a wider range of cosmologies, at scales $k<10 \ h/$Mpc and $z<2$.

Here it is worth to notice that the HACC simulations~\cite{Mira_Titan_I,Mira_Titan_II}, from which the \textsc{Mira Titan} emulator is built, do not include neutrinos as a separate species, but rather the latter are incorporated \textit{a posteriori} by adding the linear neutrino power spectrum to the nonlinear baryon-CDM component at each redshift of interest~\cite{neutrino_correction-Upadhye+14,Mira_Titan_I,Mira_Titan_II}.

The halo-model reaction uses the halo-model~\cite{HMcode2016_2,Halo_model,Cacciato:2008hm,Giocoli:2010dm} to predict corrections to the $\Lambda$CDM matter power spectrum in beyond-$\Lambda$CDM scenarios. The nonlinear power spectrum is then written as
\begin{equation}
    P^{\rm NL}(k,z) = \mathcal{R}(k,z)  \ P_{\rm \Lambda CDM}^{\rm NL} (k,z) \, , \label{eq:nlps}
\end{equation}
where $\mathcal{R}(k,z)$ is called the halo-model reaction and $P_{\rm \Lambda CDM}^{\rm NL}(k,z)$ is the nonlinear matter power spectrum in the $\Lambda$CDM case.
The method was improved upon Ref.~\cite{React_I} adapting the reaction to correct a nonlinear \textit{pseudo} power spectrum, $P_{\rm pseudo}^{\rm NL}$, rather than the $\Lambda$CDM spectrum, where the nonlinear pseudo spectrum differs from the \LCDM spectrum in that its initial conditions are tuned so that, at a target redshift, the linear clustering of the total matter matches the linear clustering in the beyond-$\Lambda$CDM cosmology.
The pseudo spectrum can be obtained using a nonlinear prescription, {\it e.g.} HMcode2020~\citep{HMcode2020}, with the $\Lambda$CDM settings but with the linear spectrum in the new cosmology as input, as implemented in this work. Note that we have made the explicit assumption that $\mathcal{R}$ and $P_{\rm \Lambda CDM}^{\rm NL}$ can be modelled independently, and that the nonlinear $\Lambda$CDM-specific physics ``drops out'' from the computation of $\mathcal{R}$. This has been shown to be a good assumption in a number of works (see Ref.~\citep{HMcode2016_2} for example).

The reaction takes the form of a ratio of halo-model predictions for the power spectrum which cancels out intrinsic halo-model inaccuracies (see Ref.~\cite{Halo_model} for a review).
In this work we follow Refs.~\cite{React_III,React_V} which provide the reaction prescription as 
\begin{equation}
    \mathcal{R}(k)=\frac{\left(1-f_{\nu}\right)^{2} P^{\rm HM}_{\rm cc}(k)+2 f_{\nu}\left(1-f_{\nu}\right) P^{\rm HM}_{\rm c\nu}(k)+f_{\nu}^{2} P^{\mathrm{L}}_{\nu\nu}(k)}{P^{\mathrm{L}}_{\mathrm{mm}}(k)+P^{\mathrm{1h}}_{\mathrm{pseudo}}(k)} \, .
    \label{eq:reaction}
\end{equation}
The effects of massive neutrinos are included~\citep{Agarwal:2010mt} at the linear level in the numerator through the weighted sum of the massive neutrino linear spectrum, $P^{\mathrm{L}}_{\nu\nu}$, and the nonlinear halo-model spectra, $P^{\rm HM}_{\rm cc}$ and $P^{\rm HM}_{\rm c\nu}$.
The components of the reaction are
\begin{equation}
    P^{\mathrm{HM}}_{\rm c\nu}(k) \approx \sqrt{P^{\mathrm{HM}}_{\rm cc}(k) P^{\mathrm{L}}_{\nu\nu}(k)} \, , 
\end{equation}
\begin{equation}
    P^{\mathrm{HM}}_{\rm cc}(k) = P^{\mathrm{L}}_{\rm cc}(k)+P^{\mathrm{1h}}_{\rm cc}(k) \, , \label{eq:1hcb} 
\end{equation}
where ``HM'' stands for halo-model, and $P^{\mathrm{1h}}_{\rm cc}$ is the 1-halo term for the CDM+b component. 

We refer the reader to Refs.~\cite{React_III,React_V} for details of the halo-model reaction in the specific case of massive neutrinos, and to Ref.~\cite{React_I} for the case of DDE. 
In this work we use the publicly available code \react\footnote{\url{https://github.com/nebblu/ReACT}} to compute the reaction $\mathcal{R}$.
By construction, the accuracy of this approach is limited to the accuracy
of the pseudo spectrum \cite{React_I,React_IV}, which in this work is chosen to be the HMcode2020 prediction.
The combination of HMcode2020 and \react \ might seem at first a double-halo model calculation. In fact, HMcode2020 has not been fitted to any emulator involving massive neutrinos and/or DDE, so that we feel safe to enhance the latter predictions in these extended cosmologies using \react.

We note that one can also use a more accurate emulator-based approaches to model the pseudo spectrum~\citep{React_II}, but this is only really feasible for modifications to $\Lambda$CDM involving only a shift in the linear power. This is because one can simply tune the amplitude parameter, $\sigma_8$ or $A_s$ say, to match the modified cosmology at linear scales and expect the appropriate pseudo cosmology modifications at nonlinear scales. For the massive neutrino cosmologies considered in this paper, which introduce a non-trivial scale dependence to the linear power, we cannot easily use such emulator-based pseudo spectra.

On the emulator side, we consider in this work both \baccoemu \ and EuclidEmulator2.
The former, presented in Ref.~\cite{baccoemu_sims} and expanded in Refs.~\cite{baccoemu_baryons,baccoemu_linear}, aims to provide accurate predictions for CDM+b power spectra in massive neutrinos and DDE cosmologies.
With only six high-resolution \Nbody simulations and thanks to the extensive use of rescaling-cosmology algorithms~\cite{rescaling_algorithm-Angulo+10,rescaling_algorithm_neutrinos-Zennaro+19}, \texttt{baccoemu}\footnote{\url{https://baccoemu.readthedocs.io/en/latest/index.html}} accurately predicts the nonlinear boost to the CDM+b power spectrum as
\begin{equation}
    B_\texttt{baccoemu}(k) = \frac{P_\mathrm{cc}^\mathrm{NL}(k)}{P_\mathrm{cc}^\mathrm{L}(k)}.
\end{equation}
The emulated boost factor is accurate at the 2\% level for the \LCDM model at scales $k\leq 5 \ h/$Mpc and redshifts $z\leq1.5$ (i.e. the maximum redshift at which \baccoemu \ is trained), while it degrades to 3\% when extended to DDE and massive neutrinos.
In this work, once $B_\texttt{baccoemu}(k)$ is obtained, $\Pmm^{\rm NL}(k)$ is computed through Eq.~\eqref{eq:power_spectrum_cdm_nu} (and treating neutrinos linearly as described above).
Moreover, the model with $M_\nu=0.32$ eV, $w_0=-0.9, w_a=0.3$ has a $\sigma_8$ value outside of the \baccoemu\ training range and will therefore be absent from the results presented here.

On the other hand, EuclidEmulator2\footnote{\url{https://github.com/miknab/EuclidEmulator2}} (version 2)~\cite{EuclidEmulator2,EuclidEmulator} is the new version of the official Euclid predictor for the boost factor in cosmologies with DDE and massive neutrinos.
Unlike \baccoemu, EuclidEmulator2 predicts the nonlinear boost factor for total matter,
\begin{equation}
    B_\mathrm{EE2}(k) = \frac{P_\mathrm{mm}^\mathrm{NL}(k)}{P_\mathrm{mm}^\mathrm{L}(k)}.
\end{equation}
The emulated nonlinear boost factor is accurate, compared to the simulations, at the 1\% level for $0.01 \leq k /(h/\mathrm{Mpc}) \leq 10$ and $z\leq 3$.

A small remark must be stated here: in its current version, the parameter space of EuclidEmulator2 extends to neutrino masses up to 0.15 eV.
We feel safe however, to use this emulator at least for our minimum non-null neutrino mass which is 0.16 eV, assuming that the error in such extrapolation is negligible. To this end, we tweak the EuclidEmulator2 code in order not to return errors for $M_\nu\leq0.16$ eV. We do not use this emulator for the models with $M_\nu=0.32$ eV.

%%%%%%%%%%%%%%%%%%%%%%%%%%%%%%%%%%%%%%%%%%%%%%%%%%%%
\section{Results}
\label{sec:results}

As a first step, we compare, at different redshifts, the overall behaviour of the various nonlinear prescriptions listed in Sec.~\ref{sec:nonlinear_methods} against the fiducial \LCDM DEMNUni data, in order to have an idea of the expected matching about models and simulated measurements also in extended cosmologies. 
Then we make a comparison of the various methods in predicting the nonlinear matter power spectra as well as the response functions, Eq.~\eqref{eq:response_neutrino_dark_energy}, in all the $\nu w_0 w_a$CDM cosmologies covered by the DEMNUni simulations.
Finally, we turn our attention to the differences in the cosmic shear and galaxy clustering angular spectra arising from using different methods to predict the nonlinear matter power spectra.
%===================================================
\subsection{Nonlinear total matter power spectra}
\label{subsec:nonlinear_spectra}
\subsubsection{$\Lambda$CDM cosmology}
%---------------------------------------------------
\begin{figure}[t]
\centering 
\includegraphics[width=0.99\textwidth]{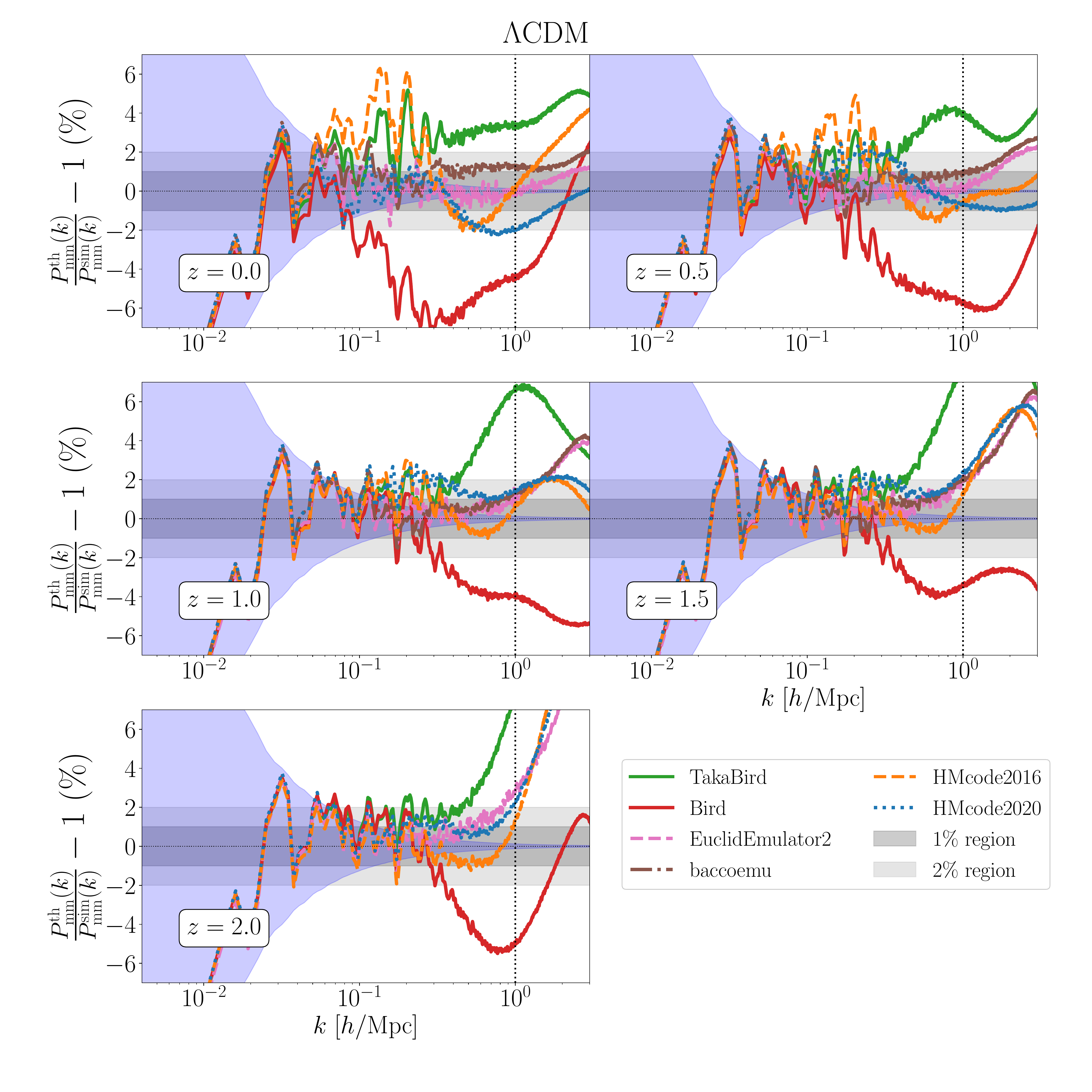}
\caption{Ratios of total matter nonlinear power spectra obtained via the models described in Sec.~\ref{sec:nonlinear_methods} with respect to DEMNUni spectra, for the \LCDM case.
The lines representing such ratios correspond to: solid green for the TakaBird model, solid red for the Bird model, dot-dashed brown for \baccoemu, dashed pink for EuclidEmulator2, dashed orange for HMcode2016 and dotted blue for HMcode2020.
Due to the limited redshift range of \baccoemu, its prediction at $z=2$ is missing; the \react~method is also missing because in \LCDM it coincides with HMcode2020 by construction.
The two grey shaded areas represent the 1\% (dark) and 2\% (light) regions, respectively. Finally, the blue shaded area shows the expected cosmic variance given the volume of the DEMNUni particle snapshots.
In this and the following plots, shot noise is added to the theoretical predictions.}
\label{fig:LCDM}
\end{figure}
%---------------------------------------------------
In Fig.~\ref{fig:LCDM} we show the ratios between the theoretical predictions of the nonlinear matter power spectrum from the various prescriptions introduced in Sec.~\ref{sec:nonlinear_methods} and the measurements from the reference \LCDM run of the DEMNUni suite.
At large scales cosmic variance dominates the uncertainty and causes significant deviations from the data, vanishing any effort to determine the accuracy of the different prescriptions in that regime.
As a consequence, hereafter, when assessing the overall performance of a nonlinear model, we will always refer to $k>k_{\rm min}\sim 5\times 10^{-2} \ h/$Mpc, i.e. to wavenumbers larger than the $k$ at which the accuracy is comparable or even smaller than the cosmic variance.
Moreover, given the precision on the matter power spectra of the DEMNUni simulations, expected to be of about $1\%$ at $k\sim 1 \ h/$Mpc and $z\lesssim 1$, as determined by their mass resolution~\cite{accuracy_power_spectrum-Schneider+16}, in the following we will consider only scales $k\lesssim 1 \ h/$Mpc (marked by the vertical dotted lines in Fig.~\ref{fig:LCDM}).
In our comparisons, shot noise is always added to the nonlinear theoretical prescriptions rather than subtracted from the measured power spectra.

In the \LCDM case, we find that:
\begin{itemize}
    \item EuclidEmulator2 (pink dashed line) best matches the simulated data. Up to $z=2$ it achieves an agreement of 2\% at $k\lesssim 1 \ h/$Mpc, with the DEMNUni spectra differing, up to $k\lesssim 3 \ h/$Mpc, by less than 1\% at $z=0$, and less than 2\% at $z=0.5$.
    \item \baccoemu \ (brown dash-dotted line) behaves similarly to EuclidEmulator2 especially at $z\gtrsim 1$. At smaller redshifts and $k\lesssim 2 \ h/$Mpc, it shows a level of agreement with the DEMNUni spectra of $\sim 1\%$  at $z=0$, and $\sim 2\%$  at $z=0.5$. A new version of this emulator, with improved concentration correction, is currently under development and is expected to further increase the accuracy of \baccoemu \ below 1\%. However, this version is not yet publicly available and, therefore, is not included in this analysis.

    \item HMcode2020 (blue dotted line) stays within 2\% accuracy up to $k = 1 \ h/$Mpc at all redshifts, but does not reach the same agreement as EuclidEmulator2 and \baccoemu \, especially at low redshifts. Interestingly, its agreement with them increases with redshift.

    \item HMcode2016 (orange dashed line) is within 2\% accuracy at $k\lesssim 1 \ h/$Mpc and $z\gtrsim 1$, but fails to correctly reproduce BAO features at lower redshifts, producing an excess of power at $k \gtrsim 0.03 \ h/$Mpc.
    This is somehow expected since the model does not correct for the late-time BAO smearing caused by nonlinear motions.

    \item The TakaBird (green solid line) method (which in the \LCDM case reduces to the original Takahashi \textit{et al.} 2012 model) has an accuracy of about 2\% up to $k\lesssim 0.5  \ h/$Mpc at $z \gtrsim 0.5$, but at $k\gtrsim 0.5 \ h/$Mpc it starts to show an excess of power of about 4--5\% with respect to the DEMNUni spectra. Moreover, at $z=0$ it exceeds 2\% accuracy mostly at all the considered scales. This behaviour is quite weird if compared to EuclidEmulator2.
    In fact, while an excess of power would be expected if we compare the DEMNUni mass resolution with the one adopted in the simulations against which the TakaBird fit was performed, certainly such an excess of power is not expected when the TakaBird model is compared against EuclidEmulator2, which has been tested against simulations of similar mass resolution as for the TakaBird model, much larger volume and accounting for sophisticated resolution correction factors and advanced initial conditions.

    \item Finally, the Bird (red solid line) model (which in the \LCDM case reduces to the Smith \textit{et al.} 2003 model) performs more poorly than the aforementioned approaches, due to the limited mass resolution of the simulations against which the fit was performed.
\end{itemize}

%---------------------------------------------------
\begin{figure}[t]
\centering 
\includegraphics[width=0.99\textwidth]{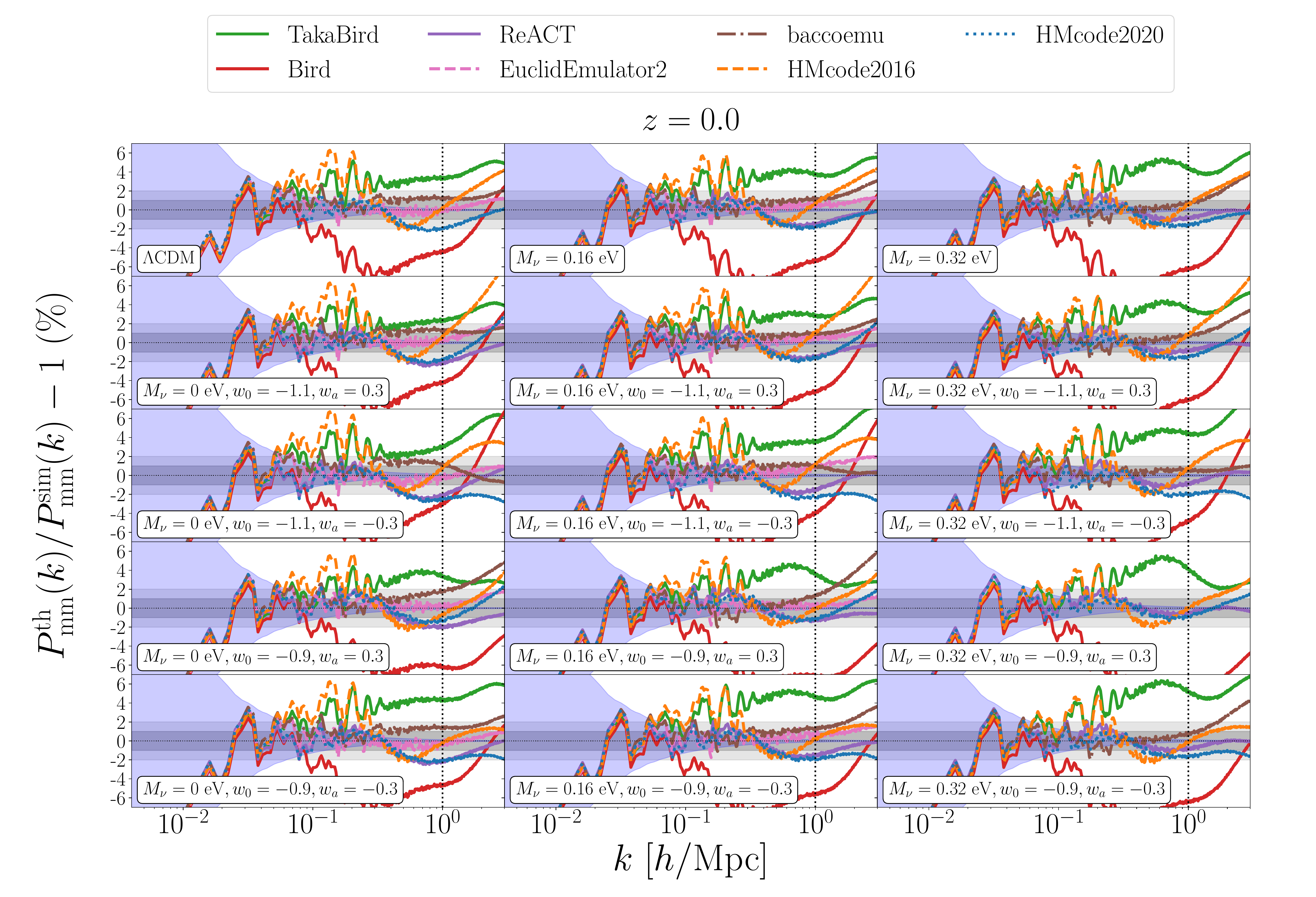}
\caption{Comparison among the various nonlinear prescriptions in predicting the nonlinear total matter power spectrum at $z=0$ for all the models considered. Solid green line: TakaBird model; solid red line: Bird model; solid purple line: \react; dot-dashed brown line: \baccoemu; dashed pink line: EuclidEmulator2; dashed orange line: HMcode2016; dotted blue line: HMcode2020. For the case $M_\nu=0.32$ eV, $w_0=-0.9$ and $w_a=0.3$, the \baccoemu~prediction is not shown as the $\sigma_8$ for this model is out of the range in which the emulator was trained. Blue shaded areas represent cosmic variance, the dark (light) grey area represent the 1\% (2\%) region.}
\label{fig:Pk_z0}
\end{figure}
%---------------------------------------------------
\begin{figure}[t]
\centering 
\includegraphics[width=0.99\textwidth]{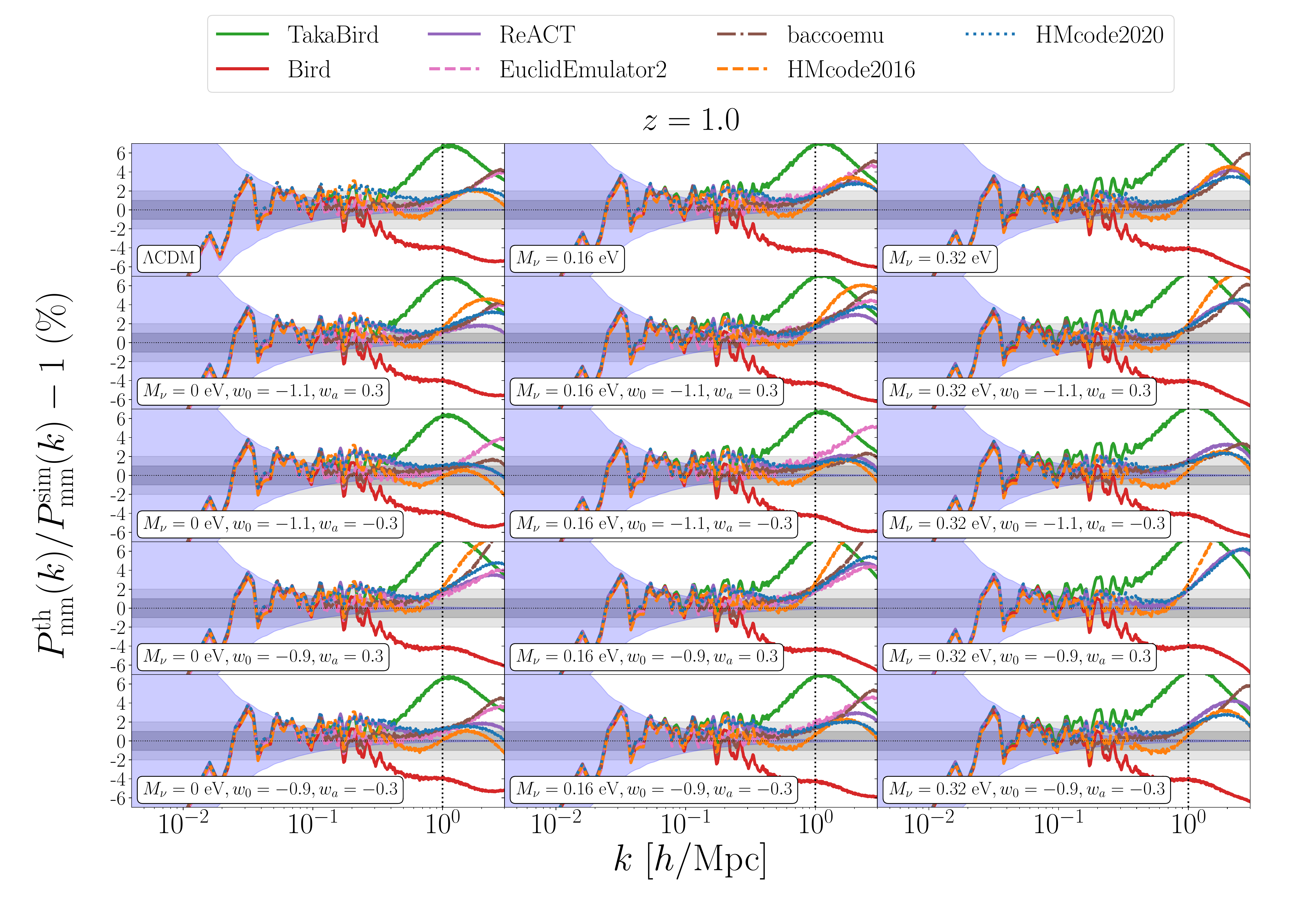}
\includegraphics[trim={0 0 0 4cm},clip,width=0.99\textwidth]{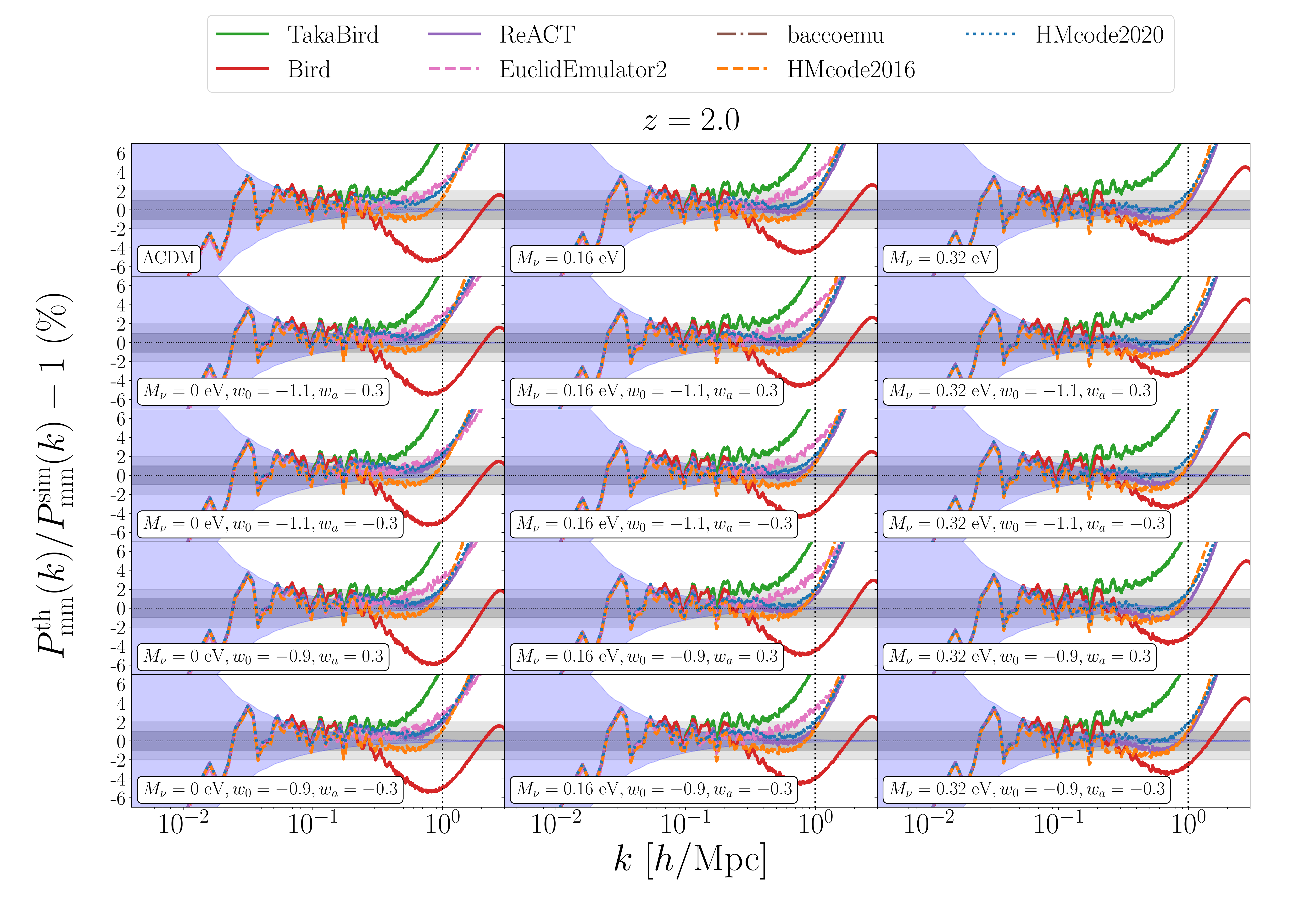}
\caption{Same of Fig.~\ref{fig:Pk_z0} but at $z=1$ (top) and $z=2$ (bottom).}
\label{fig:Pk_z1}
\end{figure}
%---------------------------------------------------
\subsubsection{$\nu w_0w_a$CDM cosmology}
While Fig.~\ref{fig:LCDM} gives hints on the performance of nonlinear prescriptions in the best-case scenario, i.e. the \LCDM\ model, we expect that the inclusion of new components, like massive neutrinos and DDE, may deteriorate this picture.
In Figs.~\ref{fig:Pk_z0} and \ref{fig:Pk_z1} we show the ratios between the matter power spectra from theoretical prescriptions and the DEMNUni power spectra, for all the models considered in this work, at $z=0$, and $z=1,2$, respectively.
We keep the same line styles and colours from Fig.~\ref{fig:LCDM} and add the solid purple line representing the prediction from \react.

The left, middle and right panels of Figs.~\ref{fig:Pk_z0} and \ref{fig:Pk_z1} correspond to total neutrino masses $M_\nu= 0,\,0.16,\,0.32$ eV, respectively.
For each cosmological model we verify the matching of the different prescriptions with the DEMNUni measurements.

Let us consider the different redshifts separately. At redshift $z=0$ Fig.~\ref{fig:Pk_z0} shows that:
\begin{itemize}
    \item for $M_\nu=0$ eV the best agreement between DEMNUni spectra and nonlinear models is again reached by the EuclidEmulator2 (pink dashed line), which is within 1\% up to $k\lesssim 3 \ h/$Mpc. For HMcode2020 (blue dotted line) and \react \ (purple solid line) the ratios corresponding to the different dark energy EoS are very similar to the \LCDM case and are within 2\% accuracy at $k\lesssim 1 \ h/$Mpc. For HMcode2016 the accuracy is similar, but worsens down to 5\% in the BAO region due to the lack of treatment of nonlinear motions on scales $<10$ Mpc$/h$.
    \baccoemu \ is always in 1--2\% agreement with the DEMNUni spectra up to $k\lesssim 2 \ h/$Mpc, though in some cases presents a large excess of power, especially for $(w_0=-0.9, w_a=0.3)$, with respect to the EuclidEmulator2 predictions. The TakaBird (green solid line) and even more the Bird (solid red line) models behave again worse than other models, with a maximum mismatch to the DEMNUni spectra of 4\% and 7\% at $k\lesssim 1 \ h/$Mpc, respectively. Nonetheless, the Takabird model seems to improve for EoS $(w_0=-1.1, w_a=0.3)$ and $(w_0=-1.1, w_a=-0.3)$.
    
    \item for $M_\nu>0$ eV all the nonlinear prescriptions improve, except for the Bird and TakaBird models which become even less accurate with increasing neutrino mass, increasing the lack of power for the former and the excess of power for the latter. We recall here that the Bird and TakaBird methods were not tailored to $w_0-w_a$ models.
\end{itemize}

At redshifts $z=1$ and $z=2$, Fig.~\ref{fig:Pk_z1} shows that:

\begin{itemize}
    \item At $z=1$ and for $M_\nu=0$ eV, except for the Bird and TakaBird models, all the nonlinear prescriptions agree within 1\% difference with the DEMNUni spectra up to $k\sim 1 \ h/$Mpc. In general, the best agreement is reached by EuclidEmulator2 and \baccoemu, which however shows again a slight excess of power with respect to the former, especially for the dark energy EoS $(w_0=-0.9,w_a=0.3)$, and except for $(w_0=-1.1, w_a=-0.3)$, where the difference between EuclidEmulator2 and \baccoemu\ is inverted with respect to the other EoS. 
    
   \item At $z=1$ and for $M_\nu>0$ eV, there is not an observable improvement, with respect to the massless neutrino case, in the agreement between the nonlinear prescriptions and the DEMNUni spectra. This is probably due to the lower impact of neutrino free-streaming at $z=1$ with respect to $z=0$. In addition, while the accuracy of the Bird model looks to stay unchanged at about $\lesssim 4\%$ for $k\lesssim 1 \ h/$Mpc as the neutrino mass increases, the TakaBird model exceeds by 7\% at $k\sim 1 \ h/$Mpc for $M_\nu=0.32$ eV\footnote{For EuclidEmulator2 we consider $M_\nu=0.16$ eV alone, as $M_\nu=0.32$ eV exceed its range of validity.}. 

    \item At $z=2$\footnote{We do not consider \baccoemu\ as $z=2$ is out of its range of validity.}, HMcode2016, HMcode2020 and \react\  agree within 1\% difference with the DEMNUni spectra up to $k\sim 1 \ h/$Mpc. EuclidEmulator2 differs from the DEMNUni spectra up to 2\% at $k\sim 1 \ h/$Mpc for $M_\nu=0$ eV. This mismatch, which increases for $M_\nu=0.16$ eV, is difficult to explain (see also the discussion in Sec.~\ref{subsec:extended_models}), unless it is associated to the linear treatment of massive neutrinos in the simulations which EuclidEmulator2 was trained on. Interestingly, the lack of power of the Bird model decreases with increasing $M_\nu$ from 6\% down to 4\% $k\lesssim 1 \ h/$Mpc, while the TakaBird model exceeds more than 7\%. 
\end{itemize}

It is worth noting that, by construction, the \react \ predictions closely follow those of HMcode2020 up to the mildly nonlinear regime, i.e. at $k \lesssim 0.1 \ h/$Mpc, but, depending on $z$, it seems that the halo-model reaction becomes  important at scales $k\sim 0.5-1 \ h/$Mpc, where the 1-halo term starts to be dominant. However, these wavenumbers are comparable to the maximum scale at which the DEMNUni simulations are expected to be accurate at about 1\% (see Sec.~\ref{sec:simulations}), so that, at the scales analysed in this work, the performance of \react \ is basically bound to the method used for the pseudo spectrum (HMcode2020 in this case) on which its predictions are based. This being said, halo model reaction predictions are expected to be $2\%$ accurate at the level of the ratio with $\Lambda$CDM, $S_{\rm mm}$. Given this, the importance of the halo model reaction cannot be properly determined in the absence of pseudo cosmology simulations.

%===================================================
\subsection{The impact of massive neutrinos and dynamical dark energy on $P^{\rm NL}_{\rm mm}$ ratios}
\label{subsec:extended_models}
%---------------------------------------------------
\begin{figure}[t]
\centering 
\includegraphics[width=0.99\textwidth]{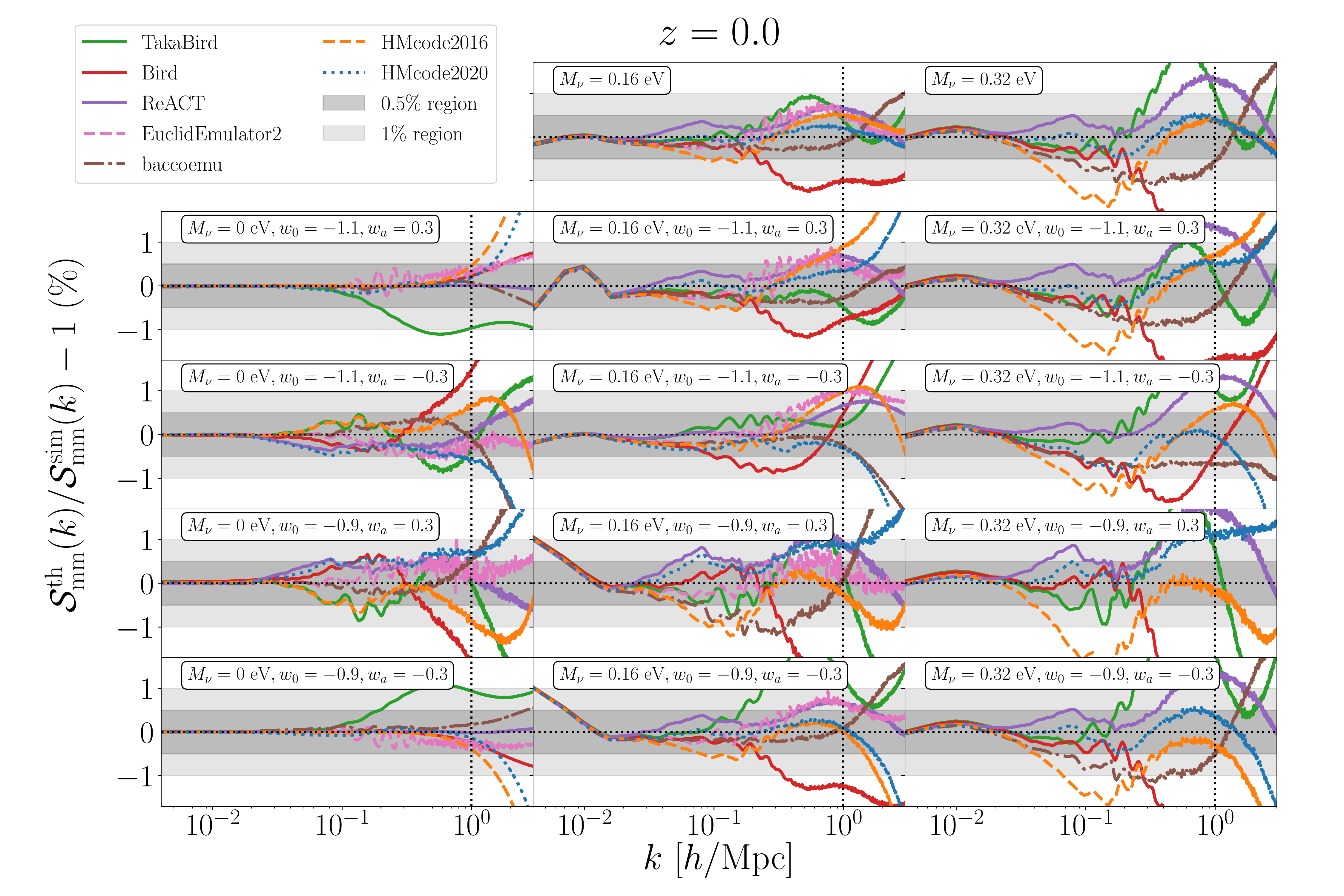}
\caption{Comparison among the various nonlinear methods in predicting the ratio of the nonlinear matter power spectrum with respect to its \LCDM counterpart, at $z=0$. Solid green line: TakaBird model; solid red line: Bird model; solid purple line: \react; dot-dashed brown line: \baccoemu; dashed pink line: EuclidEmulator2; dashed orange line: HMcode2016; dotted blue line: HMcode2020. The dark (light) grey area represents the 0.5\% (1\%) region.}
\label{fig:suppression_z0}
\end{figure}
%---------------------------------------------------
\begin{figure}[t]
\centering 
\includegraphics[width=0.99\textwidth]{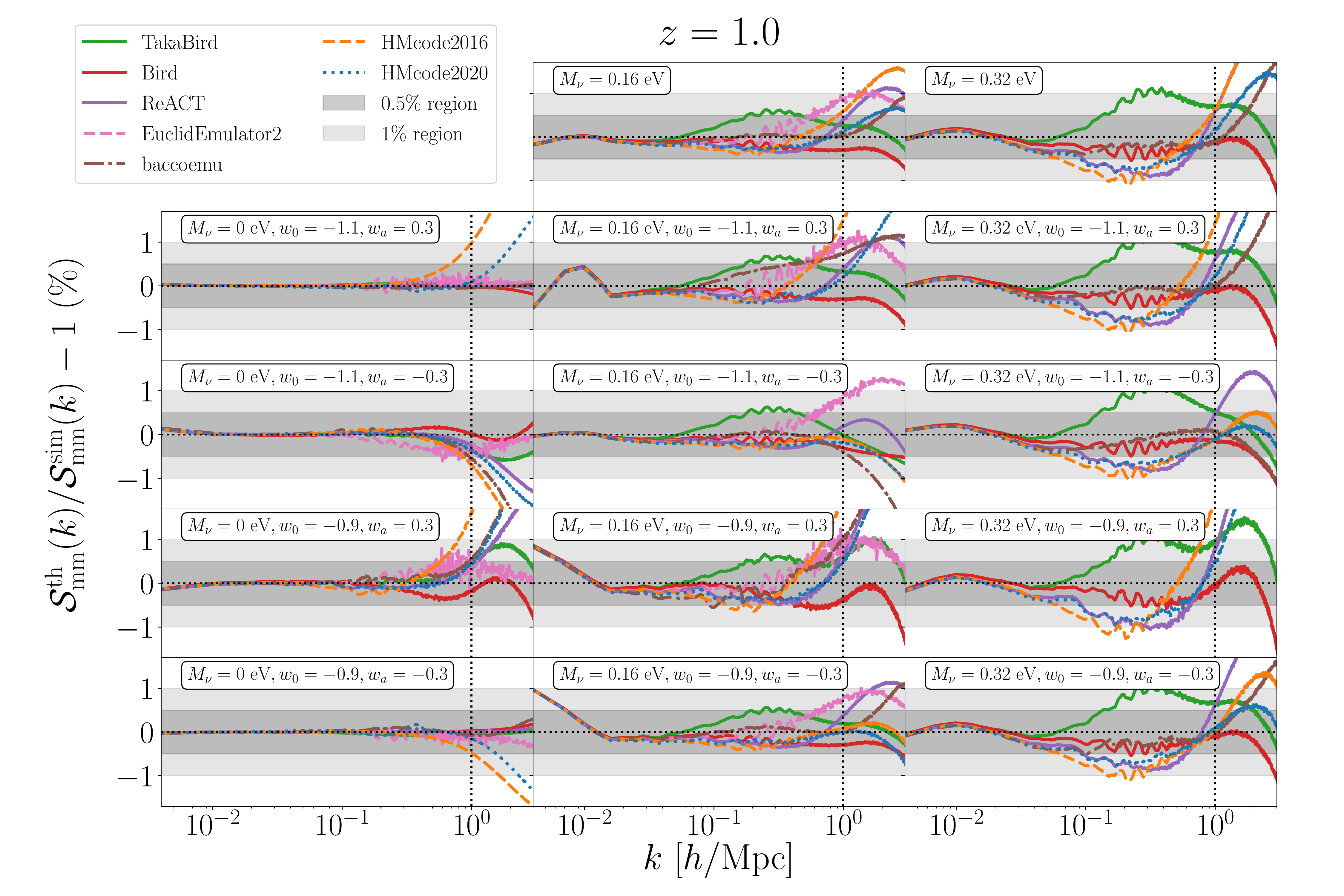}
\includegraphics[width=0.99\textwidth]{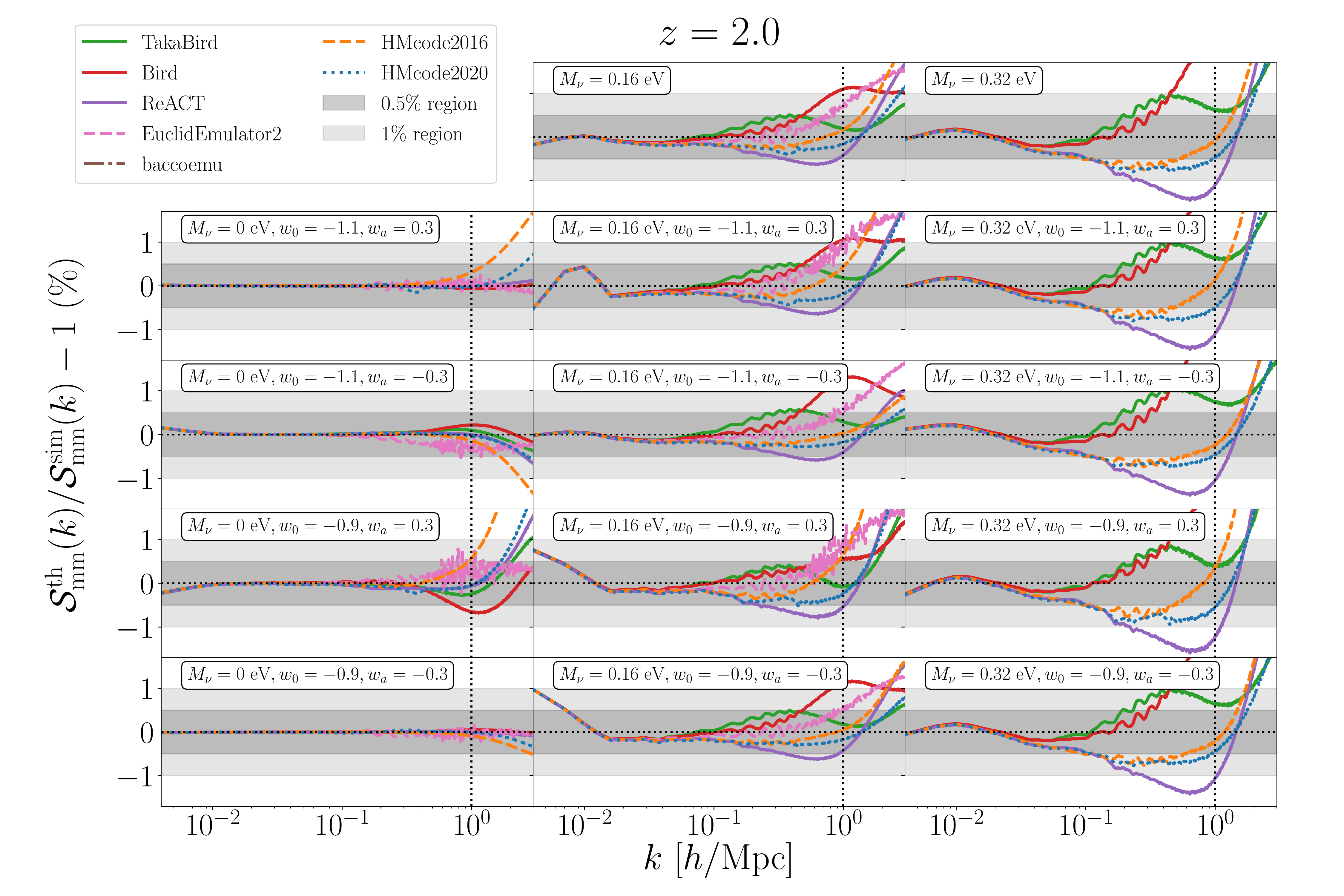}
\caption{Same of Fig.~\ref{fig:suppression_z0} but at $z=1$ (top) and $z=2$ (bottom).}
\label{fig:suppression_z1}
\end{figure}
%---------------------------------------------------
In Sec.~\ref{subsec:nonlinear_spectra} we made a comparison of the performance of different approaches in predicting the nonlinear matter power spectra in $\nu w_0 w_a$CDM cosmologies.
In this Sec. we compare their capability of predicting the ratios of nonlinear matter power spectra with respect to the \LCDM case, i.e. the response, $\mathcal S(k)$, defined in Eq.~\eqref{eq:response_neutrino_dark_energy}.
We first compute separately the neutrino mass and dark energy responses both for the nonlinear prescriptions and the power spectra measured from simulations.
Then we take the ratio between the former and the latter at $z=0$ and show the corresponding percent differences in Fig.~\ref{fig:suppression_z0}. Then, in Fig.~\ref{fig:suppression_z1} we perform a similar comparison at $z=1$ and $z=2$.
We adopt the very same structure, colours and line styles of previous Figures, while now the dark and light shaded areas represent the 0.5\% and 1\% regions, respectively.

We start focussing on the results at $z=0$ for DDE and $M_\nu=0$ eV, shown in the left panels of Fig.~\ref{fig:suppression_z0}.
In this case almost all the nonlinear models are able to predict the response to the \LCDM model with sub-percent agreement at $k \lesssim 1 \ h/$Mpc when compared to the DEMNUni response. The main exception is represented by the TakaBird model which stays within 1\% difference: it was fitted against cosmologies with a constant dark energy EoS ($w_a=0$) and therefore it is expected to break down, at least partially, in the case of DDE.
It is interesting to notice that, unexpectedly, the two cosmologies for which the TakaBird model performs more poorly are the ones for which the expansion rate, $H(z)$, is the closest to the \LCDM case (see red and green lines in Fig.~\ref{fig:effect_on_Hz}).
Instead, for all the other nonlinear models the worst case scenarios occur for the DDE models for which $H(z)$ is most different from the \LCDM\ case, in particular for the dark energy EoS $(w_0=-0.9,w_a=0.3)$ (blue line in Fig.~\ref{fig:effect_on_Hz}), where the differences between all the nonlinear prescriptions and the DEMNUni spectra are  about 0.5\%. 

Still at $z=0$ but for $M_\nu>0$ eV (middle and right columns in Fig.~\ref{fig:suppression_z0}), the nonlinear prescriptions do not achieve the same level of sub-percent agreement with the DEMNUni response as for $M_\nu=0$ eV. We think that this could be due to the scale-dependence of the neutrino-induced suppression which is more difficult to model than the different Universe background expansion associated to the DDE scenarios presented in this work.
In fact, for $M_\nu>0$ eV the ratios between the modelled and measured responses, $\mathcal{S}(k)$, become more pronounced, reaching for $M_\nu=0.16$ eV and $k \lesssim 1 \ h/$Mpc a 1\% difference in the worst case scenarios, i.e. for $(w_0=-0.9,w_a=0.3)$ and $(w_0=-1.1,w_a=-0.3)$ (blue and yellow lines in Fig.~\ref{fig:effect_on_Hz}).
Interestingly, except for $(w_0=-0.9,w_a=0.3)$, \baccoemu\ and HMcode2020 agree at sub-percent level with the DEMNUni response, while EuclidEmulator2 presents an excess of power with respect to both DEMNUni and \baccoemu\ responses.
This excess could be again due to the linear treatment of massive neutrinos in the \Nbody simulations against which EuclidEmulator2 was trained, and which could be not accurate enough for large values of the neutrino mass as $M_\nu=0.16$ eV.
For $M_\nu=0.32$ eV the differences for some nonlinear models increase beyond 1\% level, being HMcode2020 and \baccoemu\ able to stay always within 1\% agreement with the DEMNUni responses up to $k \sim 1 \ h/$Mpc.
Always at $z=0$, \react\ performs slightly worse than HMcode2020, especially at mildly nonlinear scales, even though the discrepancies with respect to the latter are still under 0.5\%. On the one hand, \react \ predictions follow closely the ones of EuclidEmulator2, with which they share a similar accuracy in most of the cases (the most discrepant is of order 0.5\% for $M_\nu=0.16$ eV, $w_0=-0.9, w_a=0.3$); on the other hand, they overpredict the suppression with respect to the DEMNUni simlations and \baccoemu, especially for $M_\nu=0.32$ eV, where the differences reach percent level.

In general, the dark energy EoS ($w_0=-0.9$, $w_a=0.3$) seems to be the most problematic, also due to the fact that, for a fixed neutrino mass, the deviation from the \LCDM spectrum is the highest $\sim10-15\%$ (see blue lines in Fig.~\ref{fig:effect_on_Pk}), and therefore more difficult to model (in case of the halo model reaction) and to include in the parameter space of emulators.

Finally, in Fig.~\ref{fig:suppression_z1}  we show the response behaviours at $z=1$ and $z=2$.
For $M_\nu=0$ eV and $M_\nu=0.16$ eV, both at $z=1$ and $z=2$, almost all the theoretical predictions remarkably agree with simulations within $\sim 0.5\%$ at $k \lesssim 1 \ h/$Mpc.
Exceptions are: \baccoemu\ which, for dark energy EoS $(w_0=-1.1, w_a=0.3)$ and $(w_0=-0.9, w_a=0.3)$,  presents a slight excess at $k \sim 1 \ h/$Mpc and $z=1$, for $M_\nu=0.16$ eV, but the same does not happen for $M_\nu=0$ eV;
EuclidEmulator2 that, for $M_\nu=0.16$ eV, presents a slight excess of response at $z=1$ and $z=2$, reaching, at $k \sim 1 \ h/$Mpc, 1\% difference with respect to the DEMNUni response. This trend is difficult to explain at such redshifts, given that it is not already present for $M_\nu=0$ eV. For $M_\nu=0.32$ eV at $z=1$ all the models, except TakaBird, agree within 1\% difference with the DEMNUni response, with \baccoemu\ performing the best, with differences below 0.5\% up to $k \sim 1 \ h/$Mpc. For $M_\nu=0.32$ eV at $z=2$, we are not able to show EuclidEmulator2 and \baccoemu\, as out from their range of validity. HMcode2016 performs slightly better than HMcode2020, both differing from the DEMNUni data by 1\% at most, while \react\ seems to underpredict the response by more than 2\%. Finally, the TakaBird model overpredicts the response by 1\% or more at all redshifts, while the Bird model, at $z=0$ underpredicts it by 2\% at $k\sim1 \ h/$Mpc, at $z=1$ performs quite well, and at $z=2$ overpredicts the response by more than 2\%.
It should be noted that {\tt ReACT} is expected to maintain a $2\%$ accuracy at the scales and redshifts considered \citep{React_V} which is consistent with out findings. Inaccuracies observed below this level cannot be attributed directly to the reaction, $\mathcal{R}$, or the pseudo spectrum, $P_{\rm pseudo}^{\rm NL}$, in the absence of pseudo cosmology simulations. The inaccuracy of HMcode2020 is also quoted as being $2.5\%$, so we cannot make any strong statements as to which of these two methods outperform the other in more general cases.

For $M_\nu=0.32$ eV at $z=1$ most models overpredict the power suppression due to neutrino free-streaming in the quasi-linear regime, despite being accurate at sub-percent level.
The best agreement is for \baccoemu.

%===================================================
\subsection{Cosmic shear and angular galaxy clustering}
\label{subsec:shear_clustering}

The matter power spectrum cannot be directly observed and therefore is accessible only through cosmological simulations.
In fact, it is involved in the theoretical modelling of several observables which are exploited for cosmological parameter inference.
Weak lensing (WL) and galaxy clustering (GC) are instead cosmological observables, and indeed two of the leading probes for constraining cosmological parameters in ongoing and upcoming surveys. In particular, WL is a direct, integrated, probe of the matter power spectrum, while GC is an indirect probe, as galaxies are biased tracers of the DM distribution. In both cases, an imprecise modelling of nonlinearities in the matter power spectrum could lead to the inference of biased best-fit parameters.

In this Sec. we investigate how well the different nonlinear prescriptions discussed in this work can predict cosmic shear and photometric galaxy clustering angular spectra measured from the DEMNUni simulations.
We limit ourselves to biases induced by the modelling of the nonlinear matter power spectra, thus neglecting other effects like intrinsic alignment (\textit{e.g.} refs.~\cite{Croft-IA+00,Heavens-IA+00,Hirata-IA+04,Troxel-IA+15}) and baryonic feedback (\textit{e.g.} refs.~\cite{HD_feedback+15,Parimbelli+19}).

The WL maps, from which we measure the angular power spectra, are obtained via ray-tracing, in the Born approximation, across the DM distribution of the DEMNUni simulations, once a full-sky lightcone has been created by means of a stacking technique, of the comoving particle snapshots, in spherical shells, with the observer placed at their centre.
This procedure follows the approaches of refs.~\citep{carbone2009, calabrese2015}, and was developed to perform high-resolution CMB and weak lensing simulations~\cite{fabbian2018,Accuracy_weak_lensing_simulations-Hilbert+20}.
In this way we produce a series of full-sky WL convergence maps on a \texttt{HEALpix}\footnote{\url{http://healpix.sourceforge.net}} grid~\citep{HEALPix} with $n_\text{side} = 4096$, which corresponds to a pixel resolution of $0.85$ arcmin. Here, we consider WL only in three different cosmological scenarios: $\Lambda$CDM and $\Lambda{\rm CDM} + M_\nu$ models with $M_\nu=0.16$ eV and $M_\nu=0.32$ eV.

The effective WL convergence map related to a source galaxy distribution can be computed as in Ref.~\citep{2001PhR...340..291B}:
\begin{equation}
    \kappa^{n(z)}_{\rm eff} (\boldsymbol{\theta}) = \int_{0}^{z_{\rm max}}  n(z) \; \kappa(\boldsymbol{\theta}, \chi(z)) \; {\rm d}z, 
\end{equation}
where $\kappa(\boldsymbol{\theta}, \chi(z))$ is the WL convergence map extracted from the DEMNUni particle lightcones at redshift $z$, and $n(z)$ is the source galaxy distribution.
The angular power spectra of these maps represent the reference \textit{simulated} signals, $C^{\rm sim}_{\gamma \gamma}(\ell)$ for the cosmic shear angular power spectra in cosmologies with massless and massive neutrino.

Concerning photometric GC, Ref.~\cite{SHAM-Carella_in_prep} has populated the DEMNUni subhalo catalogues with galaxies, via a SubHalo Abundance Matching (SHAM) method~\cite{moster_2010,girelli}, which assumes a one-to-one relation between a physical property of a dark matter halo/subhalo and an observational property of the galaxy that it hosts. As a result, applying to the SHAM catalogues a similar snapshot stacking technique as for particle snapshots, we have generated full-sky mock galaxy catalogues from the DEMNUni simulations in different cosmological models.
As in the case of the particle lightcones, we produce projected full-sky galaxy maps, representing the distribution of the SHAM galaxies on a 2D \texttt{HEALpix} grid.
The angular power spectra extracted from these galaxy maps represent the reference \textit{simulated} signals, $C^{\rm sim}_\mathrm{gg}(\ell)$, for the galaxy angular power spectra, in the considered cosmological scenarios.

From a theoretical point of view, adopting the Limber and flat-sky approximations\footnote{All the full-sky DEMNUni angular power spectra are computed on the curved sky via spherical harmonics decomposition.}, valid at large multipoles, cosmic shear and angular clustering power spectra can be written as
\begin{eqnarray}
    C_{\gamma\gamma}(\ell) &=& \int_0^{z_\mathrm{max}} \frac{\de z \ c}{H(z)} \ \frac{W^2_\gamma(z)}{\chi^2(z)} \ P_\mathrm{mm}\left(k=\frac{\ell}{\chi(z)},z\right),
    \label{eq:limber_approximation_shear}
    \\
    C_\mathrm{gg}(\ell) &=& \int_0^{z_\mathrm{max}} \frac{\de z \ c}{H(z)} \ \frac{W^2_\mathrm g(z)}{\chi^2(z)} \ P_\mathrm{cc}\left(k=\frac{\ell}{\chi(z)},z\right),
    \label{eq:limber_approximation_gg}
\end{eqnarray}

where the window functions for shear and clustering are given by
\begin{eqnarray}
    W_\gamma(z) &=& \frac{3}{2} \Omega_\mathrm m \ \frac{H_0^2}{c^2} \ \chi(z) \ (1+z) \int_z^{z_\mathrm{max}} \de z' \ n(z') \ \frac{\chi(z')-\chi(z)}{\chi(z')}, \\
    \label{eq:window_function_cosmic_shear}
    W_\mathrm g(z) &=& b(z) \ n(z) \ \frac{H(z)}{c},
    \label{eq:window_function_galaxy_clustering}
\end{eqnarray}

respectively.
In the equations above, $z_\mathrm{max}$ is set to the redshift of the lookback-time farthest snapshot used to generate the lightcones and populate the subhaloes.
The source distribution, $n(z)$, is measured directly from the DEMNUni mock galaxy lightcones, while the galaxy \textit{scale-independent} bias, $b(z)$, is computed as the square root of the ratio between the measured galaxy power spectra and  dark matter power spectra averaged on small $k$, where nonlinear effects can be neglected. To this aim, we consider comoving galaxy and DM snapshots at five redshift, $z=0,0.5,1,1.5,2$ and then interpolate in redshift the obtained values with a cubic spline.
Finally, we use the \texttt{colibri}\footnote{\url{https://colibri-cosmology.readthedocs.io/en/latest/}} software to compute the spectra in Eqs.~\eqref{eq:limber_approximation_shear}-\eqref{eq:limber_approximation_gg}, where $P_\mathrm{mm}$ and $P_\mathrm{cc}$ are computed using Eq.~\eqref{eq:power_spectrum_cdm_nu} and the nonlinear prescriptions discussed in the previous sections.
%---------------------------------------------------
\begin{figure}[t]
\centering 
\includegraphics[width=0.99\textwidth]{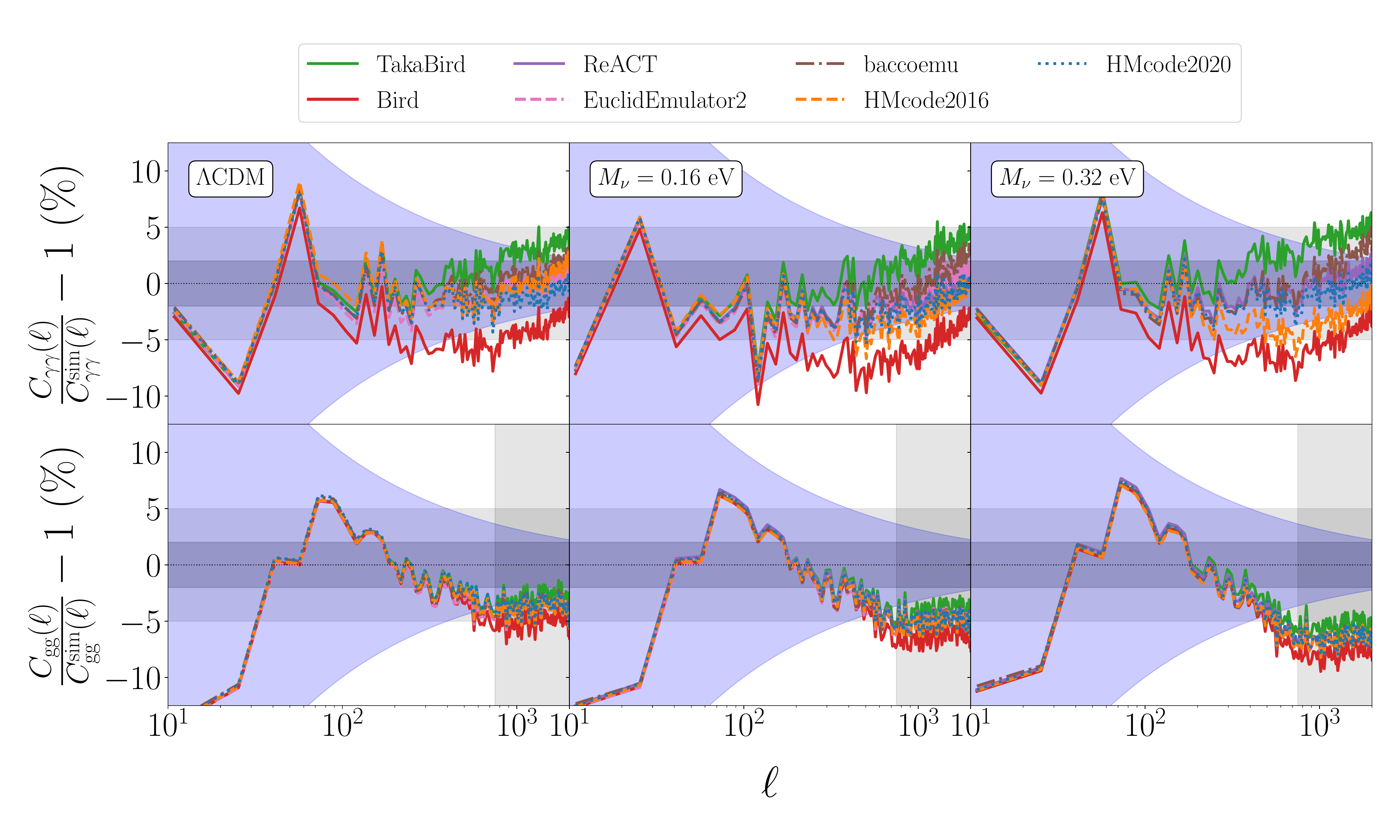}
\caption{Cosmic shear (top panels) and galaxy clustering (bottom panels) spectra comparison between different nonlinear methods and the DEMNUni simulations for three different cosmologies (fiducial \LCDM on the left, $M_\nu=0.16$ eV in the middle and $M_\nu=0.32$ eV on the right). Line styles and colours are the same of previous Figures. In all the panels, shot noise is added to the theoretical prediction. Blue shaded areas represent cosmic variance, the dark (light) grey area represents the 2\% (5\%) region. The grey vertical band in the bottom panels is an estimate of the multipole range which will not be used in upcoming surveys.}
\label{fig:shear_clustering_comparison_sims}
\end{figure}
%---------------------------------------------------

In Fig.~\ref{fig:shear_clustering_comparison_sims} we show the ratios between the predicted and simulated shear and galaxy angular power spectra.
For sake of clarity, we binned the data in bins of size $\Delta\ell =16$, weighting each data point by $2\ell+1$.
As usual, shot noise is summed to the theoretical $C(\ell)$.

As far as cosmic shear is concerned (top panels in Fig.~\ref{fig:shear_clustering_comparison_sims}), for the \LCDM model we see an overall agreement of about 2\% up to $\ell \sim 2000$ between predictions and simulations for HMcode2020, HMcode2016, \baccoemu\footnote{Given the limited redshift range of \baccoemu, for $z>1.5$ we use HMcode2020 in its place.} and EuclidEmulator2 (the \texttt{ReACT} prediction is by construction identical to HMcode2020).
On the other hand, the Bird and TakaBird methods are found to be accurate at the 5--7\% level.
The agreement remains pretty stable when including massive neutrinos: we report a slight worsening for HMcode2016 down to 5\% for $M_\nu=0.32$ eV, and for HMcode2020 down to 3\% for $M_\nu=0.16$ eV.

The situation is much more subtle for galaxy clustering (bottom panels in Fig.~\ref{fig:shear_clustering_comparison_sims}).
The shot noise increases dramatically when considering for GC the projected galaxy maps rather than for WL the DM particle maps.
Its effects are already visible at $\ell\gtrsim 300$.
This justifies the fact that at large $\ell$, where noise completely dominates the signal, the relative differences among all the nonlinear prescriptions are negligible.
We put a vertical grey band on the right of $\ell=750$, marking the pessimistic maximum multipole which Euclid will consider for galaxy clustering~\cite{Euclid_forecast+19}.
To be conservative, we chose $\ell=750$ and not the optimistic Euclid setting of $\ell=3000$: due to the minimum halo mass in the DEMNUni catalogues, the mock galaxy maps are characterised by a surface density of 8.6 galaxies per arcmin$^{-2}$, therefore by a shot noise more than three times larger than expected from the Euclid photometric sample.
Moreover, since in Eq.~\eqref{eq:limber_approximation_gg}, we use a scale-independent bias as measured from the DEMNUni galaxy and matter spectra, which holds only in the linear and mildly nonlinear regimes at most, we cannot push our analysis beyond $\ell\sim 1000$.
Overall, however, we find that all predictions provide at $\ell \gtrsim 200$ an agreement between the analytical $C_\mathrm{gg}(\ell)$ and the simulated $C^{\rm sim}_\mathrm{gg}(\ell)$ of about 3\% in the \LCDM case, and 5\% for $M_\nu=0.16, 0.32$ eV.
We believe that such a large discrepancy with respect to simulations does not quite come from GC measurements (otherwise we would see similar features for WL), but rather from the simplistic assumptions we used to compute the prediction itself. First, we assumed a linear and scale-independent galaxy bias for each snapshot, subsequently interpolating it in redshift.
Second, we assumed a Poisson shot noise.
The systematics introduced in such approximations are challenging to quantify, especially because of their possible interplay: indeed, nonlinearities in matter fluctuations and in galaxy bias typically become relevant already at $\ell\sim200-300$, in the same range of multipoles where shot noise starts to dominate.
%---------------------------------------------------
\begin{figure}[t]
\centering 
\includegraphics[width=0.99\textwidth]{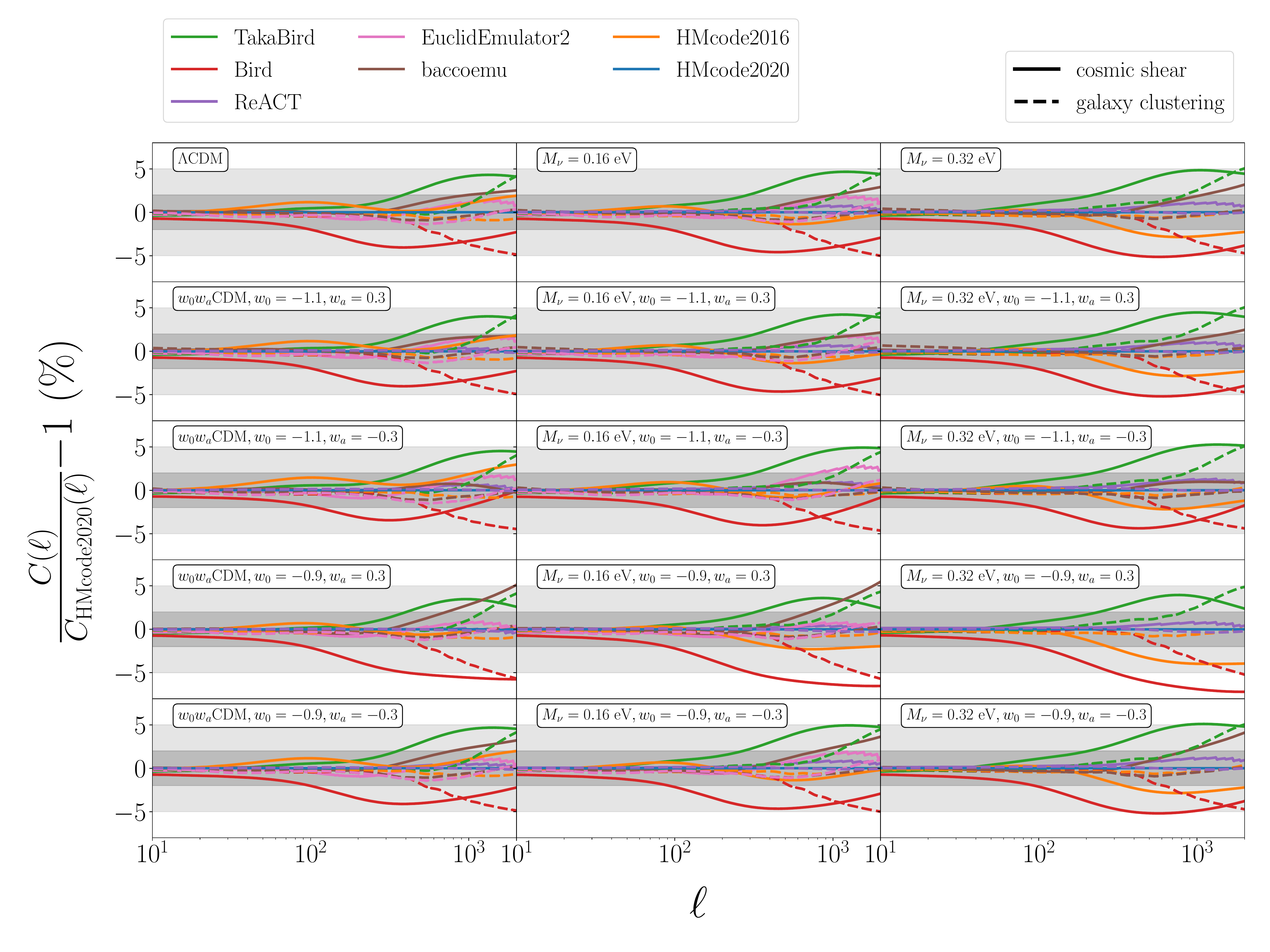}
\caption{Different prediction for cosmic shear (solid lines) and angular galaxy clustering (dashed lines) power spectra arising from the use of different nonlinear methods. 
We show here the ratio with the prediction from HMcode2020 (blue line)
Green line: TakaBird model; red line: Bird model; pink line: EuclidEmulator2; purple line: \texttt{ReACT}; brown line: \texttt{baccoemu}; orange line: HMcode2016. The dark (light) grey area represents the 2\% (5\%) region.
Due to its limited range in redshift, for $z>1.5$ we use HMcode2020 in place of \texttt{baccoemu}.}
\label{fig:shear_clustering_comparison}
\end{figure}
%---------------------------------------------------

To emphasize the differences among the $C(\ell)$ predicted from nonlinear methods, and to have an idea of their behaviour in cosmologies with DDE, we present Fig.~\ref{fig:shear_clustering_comparison}.
In each panel, solid lines represent the ratio between the cosmic shear power spectra computed using different prescriptions with respect to spectrum computed via a reference prescription. Analogous ratios are shown for the GC spectra as dashed lines.
We keep the same colours as in previous Figures to label the different nonlinear prescriptions, picking HMcode2020 as the reference model.
Our results show that almost all the models agree within $\sim5\%$.
The only exception is the case of cosmic shear for $M_\nu=0.32$ eV, $w_0=-0.9, w_a=0.3$, where the Bird model deviates at most by $\sim 7\%$ at multipoles of $\ell\gtrsim10^3$.
The TakaBird model follows a similar trend but on the opposite side, overpredicting by $\sim 5\%$ the signal in the same range of multipoles.
The remaining nonlinear methods, for both cosmic shear and galaxy clustering, all fall within the 2\% agreement, which corresponds to the typical accuracy between different lensing codes for multipoles $\ell\lesssim 4000$~\cite{Accuracy_weak_lensing_simulations-Hilbert+20}.

An important message brought by Fig.~\ref{fig:shear_clustering_comparison} is the following: all the prescriptions discussed in this work exhibit a similar behaviour regardless of the underlying cosmological model.
This means that most of the modelling uncertainty is intrinsic to the nonlinear method itself and that possible biases that may arise due to the modelling are independent of neutrino mass and dark energy.
The systematics arising from a wrong modelling of nonlinearities must be taken into account when performing cosmological inferences: while for DES they have been shown to be under control \cite{DES_yr3_extension_to_LCDM}, it might not be true for denser and/or deeper surveys like Euclid. 
Ref.~\cite{Impact_non-linear_recipe-Martinelli+21} already pushed the analysis in this direction, by quantifying the biases on the posterior of the cosmological parameters induced by a wrong choice of nonlinear modelling and baryonic feedback.
However, their fiducial model consisted of synthetic data, generated using the TakaBird prescription.
We plan to tackle this problem, using data directly from simulations, in a future work.

%%%%%%%%%%%%%%%%%%%%%%%%%%%%%%%%%%%%%%%%%%%%%%%%%%%%
\section{Conclusions}
\label{sec:conclusions}

In this work we tested how different prescriptions of the matter power spectrum in the nonlinear regime compare to \Nbody simulations in massive neutrino and dynamical dark energy cosmologies.
The simulations we employ, the DEMNUni suite~\cite{DEMNUni_simulations}, are state-of-the-art in the treatment of neutrinos as a particle species separate from CDM+b and, given their mass resolution and volume, are expected to be accurate at 1\% percent level up to wavenumbers of $k\sim 1 \ h$/Mpc~\cite{accuracy_power_spectrum-Schneider+16}.
The methods we tested fall into three different classes: fitting functions, halo-model and emulators.
In the first category, in which usually the matter power spectrum functional form is based on the halo-model, phenomenological or physically-motivated functions are calibrated against the matter power spectrum measured from simulations.
The HALOFIT models by Smith~\cite{HALOFIT_Smith} and Takahashi~\cite{HALOFIT_Takahashi} (both combined with the Bird correction to account for massive neutrino effects~\cite{HALOFIT_Bird}), HMcode2016~\cite{HMcode2016_1,HMcode2016_2} and HMcode2020~\cite{HMcode2020} belong to this first category.
In the halo-model class we test the halo-model reaction method (first presented in Ref.~\cite{React_I}).
In this approach the power spectrum in a non-\LCDM cosmology is given by a pseudo spectrum multiplied by a reaction term that involves ratios of power spectra computed via the halo-model, so that possible inaccuracies of the latter cancel out.
Finally, we use the two Euclid~\cite{EuclidEmulator2} and \baccoemu~\cite{baccoemu_sims} emulators as representative methods of the third class mentioned above.

In the (massless neutrino) \LCDM cosmology (see Fig.~\ref{fig:LCDM}) we find that, especially at low redshifts, EuclidEmulator2 best matches the simulated data at $\lesssim 1\%$ level even for $k > 1 \ h$/Mpc, reaching a maximum difference of $\sim 2\%$ at $z=2$. \baccoemu\ behaves similarly to EuclidEmulator2 especially at $z \gtrsim 1$. Overall, these emulators, together with HMcode2020, achieve a 2\% agreement with the DEMNUni spectra up to $z=2$ and $k\lesssim 1 \ h$/Mpc. 

In $\nu w_0 w_a$CDM cosmologies, at $z=0$ (see Fig.~\ref{fig:Pk_z0}), we find that, for $M_\nu=0$ eV, the best agreement between DEMNUni spectra and nonlinear models is again reached by EuclidEmulator2 which is within 1\% difference up to $k \lesssim 3 \ h$/Mpc. In this case, \baccoemu\ has an agreement of 1--2\% up to $k \lesssim 2 \ h$/Mpc, while HMcode2020 and \react\ are within 2\% accuracy at $k \lesssim 1 \ h$/Mpc. For $M_\nu>0$ eV, all the nonlinear prescriptions improve, except for the Bird and TakaBird models which, however, are not tailored to $w_0$-$w_a$ models. At $z>0$ (see Fig.~\ref{fig:Pk_z1}), we find that, for $M_\nu=0$ eV, excluding again the Bird and TakaBird models, all the nonlinear prescriptions agree within $\sim 1\%$ difference with the DEMNUni spectra up to $k \sim 1 \ h$/Mpc, but at $z=2$ EuclidEmulator2 differs by $\lesssim 2\%$. For $M_\nu>0$ eV, there is not an observable improvement, with respect to the
massless neutrino case, probably due to the lower impact of neutrino free-streaming at redshifts larger than zero; rather at $z=2$ EuclidEmulator2 exceedes $2\%$ agreement for some dark energy EoS.

In general, HMcode2016 has a similar behaviour to HMcode2020 but at low redshifts its accuracy worsens to $\sim 5\%$ at mildly nonlinear scales, due to a missing nonlinear treatment of the BAO wiggles. On the other hand, the Bird and TakaBird models are in general accurate at $\sim5\%$ but in different directions: the latter overpredicts and the former underpredicts the total matter power spectra both with respect to the other prescriptions and to the DEMNuni data.

We also investigate the precision with which the considered nonlinear prescriptions are able to predict the power spectrum response $\mathcal S(k)$, Eq.~\eqref{eq:response_neutrino_dark_energy}, i.e. the ratio between the power spectrum in a given cosmology and the corresponding \LCDM one.
We find (see Figs.~\ref{fig:suppression_z0}-\ref{fig:suppression_z1}) that the vast majority of the nonlinear models are accurate at better than percent level, even compared to responses as high as $\sim 30\%$, especially for high neutrino masses.
We confirm the same trend of the analysis for the full power spectrum: EuclidEmulator2, \baccoemu\ and HMcode2020 provide the best agreement with the DEMNUni data, while the Bird and TakaBird versions of HALOFIT are less accurate, especially at low redshifts and for DDE models, for which however, they have not been tailored.

Finally, we examine how the choice of a given nonlinear prescription affects the predictions on the cosmic shear and angular galaxy clustering power spectra.
We find (see Fig.~\ref{fig:shear_clustering_comparison_sims}) that for cosmic shear HMcode2020, \texttt{baccoemu}, EuclidEmulator2 and \texttt{ReACT} agree with the simulated $C^{\rm sim}_{\gamma \gamma}(\ell)$ by 2--3\% level at multipoles $100\lesssim \ell \lesssim2000$ regardless of the neutrino mass and DDE; HMcode2016 worsens from 2\% to 5\% for $M_\nu=0.32$ eV, while the Bird and TakaBird models underpredict and overpredict the $C^{\rm sim}_{\gamma \gamma}(\ell)$ spectra by 5\%, respectively.
For galaxy clustering, due to the DEMNUni minimum halo mass, the simulated signal $C^{\rm sim}_\mathrm{gg}(\ell)$ is strongly affected by shot noise at relatively low multipoles ($\ell\gtrsim300$), making it somewhat difficult to distinguish the differences between the various nonlinear prescriptions. However, the relative differences among them remain in line with those of cosmic shear.

Interestingly, the relative discrepancies among the different prescriptions do not depend dramatically on the underlying cosmological model (see Fig.~\ref{fig:shear_clustering_comparison}), so that we can conclude that the systematics introduced by choosing a particular nonlinear model are independent of the cosmological models.
However, it is important to take into account such discrepancies especially in cosmological inference analyses, as they may result in significant biases on the parameter posteriors. We plan to tackle this issue more quantitatively in future work.

%======================================================================
%======================================================================
%======================================================================

%============= ACKNOWLEDGMENTS =============
\acknowledgments
GP is supported by the INFN INDARK PD51 grant and the ``Unveiling Dark Matter and missing baryons in the high-energy sky'' grant funded by the agreement ASI-INAF n. 2017-14-H.0.
BB was supported by a UK Research and Innovation Stephen Hawking Fellowship (EP/W005654/1) and Swiss National Science Foundation (SNSF) Professorship grant (No.~170547 \& 202671). 
MC carried out part of this work while supported by a ``Research and Education'' grant from Fondazione CRT.
The OAVdA is managed by the Fondazione Cl\'ement Fillietroz-ONLUS, which is supported by the Regional Government of the Aosta Valley, the Town Municipality of Nus and the ``Unité des Communes vald\^otaines Mont-\'Emilius''.
The DEMNUni simulations were carried out in the framework of ``The Dark Energy and Massive-Neutrino Universe" project, using the Tier-0 IBM BG/Q Fermi machine and the Tier-0 Intel OmniPath Cluster Marconi-A1 of the Centro Interuniversitario del Nord-Est per il Calcolo Elettronico (CINECA). We acknowledge a generous CPU and storage allocation by the Italian Super-Computing Resource Allocation (ISCRA) as well as from the coordination of the ``Accordo Quadro MoU per lo svolgimento di attività congiunta di ricerca Nuove frontiere in Astrofisica: HPC e Data Exploration di nuova generazione'', together with storage from INFN-CNAF and INAF-IA2.

For the purpose of open access, the author has applied a Creative Commons Attribution (CC BY) licence to any Author Accepted Manuscript version arising from this submission.

%===========================================

%============= Data availability =============

\section*{Data Availability}

Supporting research data are available on reasonable request from the corresponding author.

%===========================================

%============== BIBLIOGRAPHY ===============
\bibliographystyle{JHEP}
\bibliography{bibtex}
%===========================================

\end{document}